\documentclass{article}
\usepackage{amssymb}

\input{tcilatex}
\begin{document}

\begin{center}
{\Large Unitarily manipulating in time and space a Gaussian wave-packet
motional state of a single atom in a quadratic potential field\bigskip }

Xijia Miao

Somerville, Massachusetts

Date: July 2007\bigskip

Abstract\\[0pt]
\end{center}

The paper first discusses theoretically the off-resonance selective
excitation method that is dependent on the atomic internal states and used
to generate approximately a standard coherent state of harmonic oscillator.
The coherent average method then is proposed to construct the
state-selective trigger pulse. A state-selective trigger pulse can keep
Gaussian shape unchanged but change in an internal-state-dependent form the
center-of-mass position and/or momentum of an atomic Gaussian wave-packet
motional state. A Gaussian wave-packet state is one of the simplest
wave-packet states that can be easily manipulated and controlled in time and
space. The paper also investigates how to manipulate in time and space an
atomic Gaussian wave-packet motional state by a generalized quadratic
potential field. A general quadratic Hamiltonian can affect not only the
center-of-mass position and momentum but also the complex linewidth of a
Gaussian wave-packet motional state while keep Gaussian shape of the
motional state unchanged. It is shown that generally quadratic terms of a
quadratic Hamiltonian can control directly the complex linewidth, while
linear terms of a quadratic Hamiltonian can affect only the center-of-mass
position and momentum of a Gaussian wave-packet motional state. \newline
\newline
\newline
{\large 1. Introduction }

In the paper [1] a particle picture has been used to describe intuitively
how the halting-qubit atom evolves in the state-locking pulse field and how
to construct a quantum control process to simulate the reversible and
unitary halting protocol that is insensitive to its input state, although
the quantum mechanical wave-packet picture has also used extensively in that
paper. A particle picture is particularly intuitive to describe the
decelerating and accelerating processes of a free atom and the elastic
collision process for an atom bouncing off a hard potential wall. The
correctness of a particle picture is based on the fact that a particle
picture is very close to a wave-packet picture in quantum mechanics [2].
However, a quantitative and exact calculation in quantum mechanics for the
time evolution process of a quantum system such as an atom does not use the
classical particle picture. One must use wave-packet states or more
generally quantum states to calculate the time evolution process of the atom
in quantum mechanics. The wave-packet states were used by Schr\"{o}dinger,
Dirac, and others to describe the quantum mechanical behavior of a particle
date back to the early time of the wave mechanics. A very familiar example
is that a free particle such as an atom may be described exactly by a
Gaussian wave packet state in quantum mechanics [2]. Wave-packet states and
especially Gaussian wave-packet states have been used frequently to describe
the quantum collision and scattering processes in the atomic and molecular
systems [2, 3, 4]. It can be seen in Refs. [5, 6, 7] that there is a more
extensive application of the Gaussian wave-packet states to describe a
variety of quantum dynamical processes of the atomic and molecular systems.
Though quantum mechanically it is not limited to use the wave-packet states
to describe the quantum control process of the reversible and unitary
halting protocol and to investigate the mechanism of the state-locking pulse
field [1], the quantum mechanical wave-packet states not only provide an
intuitive picture for understanding the mechanisms of the reversible and
unitary state-insensitive halting protocol and the state-locking pulse field
but also simplify the quantitative and exact calculation for the quantum
control process in the physical system of the halting-qubit atom. According
to the reversible and unitary state-insensitive halting protocol [1] the
wave-packet motional state of the halting-qubit atom should have a small
spread, the wave-packet amplitude of the motional state decays quickly with
the deviation from the wave-packet center and it is close to zero outside
the effective spread. Obviously, such a wave-packet picture is close to a
particle picture, which is just required by the reversible and unitary
state-insensitive halting protocol. A Gaussian wave-packet state satisfies
this requirement very much. It can have a very narrow linewidth or
wave-packet spread and its amplitude at a position deviating from the
center-of-mass position decays rapidly and exponentially with square of
distance between the position and the center-of-mass position. A\ Gaussian
wave-packet state is described completely by the three basic parameters: the
center-of-mass position, the mean momentum, and the complex linewidth [2, 5,
6, 7, 8]. Then a Gaussian wave-packet state is simple and easy to be treated
in theory and it is also easily manipulated and controlled in experiment. On
the other hand, it is well known that the ground state of a harmonic
oscillator is a Gaussian wave-packet state [2], while at the initial time of
the quantum control process the halting-qubit atom is prepared to be in the
ground state of harmonic oscillator [1]. Thus, it is convenient and natural
to choose the Gaussian wave-packet state to describe and calculate
quantitatively the quantum control process. Due to the fact that a Gaussian
wave-packet state has these advantages it should be better to keep Gaussian
shape of the motional state of the halting-qubit atom unchanged in the whole
quantum control process.

Unitary manipulation in time and space in a quantum system plays a key role
in implementing the reversible and unitary state-insensitive halting
protocol and realizing the efficient quantum search process [1, 12]. It is
known that the quantum control process to realize the reversible and unitary
state-insensitive halting protocol is a unitary evolution process in time
and space [1]. One of the key components of the quantum control process is
the time- and space-compressing processes which are realized by the unitary
decelerating and accelerating processes of the halting-qubit atom. It has
been shown [8] that the unitary decelerating and accelerating processes for
a free particle moving in space can be realized by the stimulated Raman
adiabatic passage (STIRAP) method [9]. In the ideal or near ideal adiabatic
condition the STIRAP\ pulse sequence can transfer completely one Gaussian
wave-packet motional state of the free atom to another [8]. It is known that
the unitary propagator of a quadratic Hamiltonian can generally keep
Gaussian shape unchanged for a Gaussian wave-packet state when the state is
acted on by the unitary propagator [5, 6, 13, 14]. The Hamiltonian of the
atom in the presence of the STIRAP\ pulse sequence is completely different
from a conventional quadratic Hamiltonian. But it is surprising that the
unitary propagator corresponding to the Hamiltonian of the atom in the
presence of the STIRAP pulse sequence does not yet change Gaussian shape of
an atomic Gaussian wave-packet motional state when the atom is irradiated by
the STIRAP pulse sequence in the ideal or near ideal adiabatic condition.
The STIRAP-based unitary decelerating and accelerating processes can
manipulate the center-of-mass position and momentum of a Gaussian
wave-packet state of a free atom in time and space, but it generally does
not control the complex linewidth of the Gaussian wave-packet state because
the imaginary part of the complex linewidth always increases linearly with
the time period of the decelerating or accelerating process. One advantage
to use the laser light fields such as the STIRAP\ pulse sequence to
manipulate an atomic Gaussian wave-packet motional state is that the
space-selective manipulation of the atom can be implemented easily. On the
other hand, a general quadratic Hamiltonian can be used to manipulate not
only the complex linewidth but also the center-of-mass position and momentum
of a Gaussian wave-packet state. This will be investigated in detail in the
paper.

This paper is devoted to the construction of the state-selective trigger
pulse and the unitary manipulation of a Gaussian wave-packet state in time
and space by using a general quadratic Hamiltonian. A state-selective
trigger pulse is a key component to realize both the reversible and unitary
state-insensitive halting protocol and the efficient quantum search process
based on the unitary quantum dynamics in time and space [1, 12]. It is
generally related to unitary manipulation and control in time and space of
the atomic center-of-mass and internal motions as well as the coupling of
the two motions. In the quantum control process the state-selective trigger
pulse is used to transfer the ground motional state of the halting-qubit
atom in the harmonic potential field to a standard coherent state [10, 11]
with a higher motional energy. It is known that a standard coherent state of
harmonic oscillator is a Gaussian wave-packet state [13]. There is a
requirement that the coherent-state excitation process induced by the
state-selective trigger pulse be dependent on the atomic internal state and
keep Gaussian shape of the atomic motional state unchanged. One of the
convenient methods to construct the state-selective trigger pulse could be
that the Hamiltonian of the halting-qubit atom in the presence of
state-selective trigger pulse is prepared to be a state-dependent quadratic
Hamiltonian. A general quadratic Hamiltonian of an atomic system may be
generated either by an external static electric or magnetic field or by the
externally applied electromagnetic wave field. An external static electric
or magnetic field usually does not induce the transitions of the atomic
internal electronic or nuclear spin states. Therefore, these uniform static
electric and magnetic fields are not suited to build up the
internal-state-selective trigger pulse. Generally, the construction of the
state-selective trigger pulse needs to use the external electromagnetic wave
field.\newline
\newline
\newline
{\large 2. The constructions for the state-selective trigger pulse}

The state-selective trigger pulse is one of the key components to construct
the quantum control process to simulate the reversible and unitary
state-insensitive halting protocol. It is used to control the halting
operation in the reversible and unitary state-insensitive halting protocol.
The properties for the state-selective trigger pulse have been described in
detail in the paper [1]. There are two basic properties for the
state-selective trigger pulse. One of which is that the trigger pulse is
internal-state-selective. Only when the halting-qubit atom is in the given
internal state can the state-selective trigger pulse take a real action on
the atom and excite the ground motional state of the atom to the coherent
state. Another is that the state-selective trigger pulse can keep Gaussian
shape unchanged for the atomic motional state. Under the state-selective
trigger pulse the initial motional state of the atom could be converted into
a standard coherent state of harmonic oscillator with a higher motional
energy. In the quantum control process the initial motional state generally
is prepared to be the ground state of the harmonic oscillator --- the
halting-qubit atom in the left-hand harmonic potential well [1]. It is well
known that the ground state of a harmonic oscillator is a Gaussian
wave-packet state [2] and so is a standard coherent state of the harmonic
oscillator [13]. Therefore, it is clear that the state-selective trigger
pulse should generate a unitary propagator for the halting-qubit atom such
that the unitary propagator does not change the Gaussian wave-packet shape
of the motional state of the halting-qubit atom. According to these
properties one may construct explicitly such a state-selective trigger pulse
for the reversible and unitary halting protocol in the physical system of
the halting-qubit atom in the harmonic potential well. One could choose a
sequence of laser light pulses as the state-selective trigger pulse in the
atomic physical system. But there could also be other schemes to construct
the state-selective trigger pulse.\newline
\newline
{\large 2.1. The non-internal-state-selective excitation process}

Before the internal-state-selective excitation process of an atom in a
harmonic potential field is discussed in next subsections, this subsection
is devoted to the investigation for the non-internal-state-selective
excitation process of the atom. If there is not the internal-state-selective
requirement, it is usually easy to excite a coherent state from the ground
state of a harmonic oscillator. In principle one of the simplest methods to
generate non-selectively a coherent state from the ground state of a
harmonic oscillator is to apply directly an external classical electric or
magnetic field to the harmonic oscillator, so that an extra force generated
by the external electric or magnetic field is exerted on the harmonic
oscillator. Then it can prove [14, 15, 16, 17] that this extra force can
drive the atom from the ground state to the coherent state with a higher
motional energy. For example, for a trapped atomic ion in the harmonic
potential well one may apply a spatially uniform driving electric field to
the trapped atomic ion. The experiments to confirm the simple method have
been carried out in the trapped ion systems [18, 19, 20]. Since the
classical driving field does not excite any internal state of the trapped
atom in the harmonic potential well and the trapped atom in any internal
state can be excited from the ground motional state to the coherent state by
the external driving field, this method is an internal-state-independent
excitation method to generate the coherent state. The coherent-state
excitation process may be described simply by the Hamiltonian of the atom in
the harmonic potential well and in the presence of the external driving
field, 
\begin{equation}
H(t)=H_{0}+H(r)+H_{1}(t),  \tag{1}
\end{equation}%
where the Hamiltonian $H_{0}$ describes the center-of-mass motion of the
atom in the harmonic potential well and in the absence of the external
driving field, 
\begin{equation}
H_{0}=\frac{p^{2}}{2m}+\frac{1}{2}m\omega ^{2}x^{2},  \tag{2}
\end{equation}%
here the oscillatory frequency $\omega =\omega (t)$ of the harmonic
oscillator may be time-dependent or may not, the term $H(r)$\ is the
internal Hamiltonian of the atom which describes the internal electronic (or
nuclear spin) motion of the atom, and $H_{1}(t)$ is the interaction between
the atom in the harmonic potential well and the external driving field. One
may not consider the internal Hamiltonian $H(r)$ of the atom in the
internal-state-independent excitation process, since the excitation process
is not dependent on any internal state of the atom. One of the simplest
interactions is $H_{1}(t)=f(t)x$ [14, 15, 16, 17]. It indicates that there
is the force exerted on the harmonic oscillator to generate the coherent
state. This linear interaction $H_{1}(t)$ may be generated by simply
applying the driving electric or magnetic field to the harmonic potential
well in a similar way to generating the harmonic potential $m\omega
^{2}x^{2}/2.$ For example, if the atom is a charged atomic ion with charge $%
q $, the interaction may be written as $H_{1}(t)=-qxE(t)$ for the spatially
uniform and time-dependent driving electric field $E(t)$ and hence the
forced function $f(t)$ is given by $f(t)=-qE(t).$ The atom is usually in
some hyperfine ground electronic (or nuclear spin) state before the external
driving electric field is applied. However, when the atom is irradiated by
the suitable external electromagnetic field, it may jump to an atomic
excited state or another hyperfine ground state. These internal electronic
(or nuclear spin) state transitions of the atom have their own transition
frequencies. Suppose that the minimum transition frequency among these
internal-state transition frequencies is much greater than the oscillatory
frequency of the external driving electric field $E(t)$. Then it is
impossible for the external driving electric field to induce the atom to
make a transition between the atomic internal states. During the external
driving field the atom stays in its initial internal state. Thus, the
coherent-state excitation process of the atom under the external driving
electric field is independent of the atomic internal states and hence it is
internal-state-independent. On the other hand, while the coherent-state
excitation process is independent of any atomic internal state, the driving
electric field $E(t)$ must be able to induce effectively the transitions
between the vibrational energy levels of the atom in the harmonic potential
well. The vibrational energy levels generally have a much smaller
energy-level space than the atomic internal energy levels. If now the
oscillatory frequency of the external electric field is set to be near the
oscillatory frequency of the harmonic oscillator --- the halting-qubit atom
in the harmonic potential well, then the external driving electric field may
induce the atom to make a transition between the vibrational energy levels,
while the atomic internal states are not affected by the driving electric
field. Since any internal state of the atom remains unchanged during the
coherent-state excitation process by the external driving electric field,
the atomic internal Hamiltonian $H(r)$ may be omitted from the total
Hamiltonian $H(t)$ of Eq. (1). Then in this case the Hamiltonian to describe
the time evolution process of the atom in the harmonic potential well and in
the presence of the external driving electric field is reduced to the form 
\begin{equation}
H(t)=\frac{p^{2}}{2m}+\frac{1}{2}m\omega ^{2}x^{2}+f(t)x.  \tag{3}
\end{equation}%
This is the Hamiltonian of a time-dependent forced harmonic oscillator [13,
14, 15, 16, 17, 21]. If now the initial motional state of the atom is the
ground state of the harmonic oscillator, then the unitary propagator of the
Hamiltonian $H(t)$ (3)\ will convert the ground state into a coherent state
[17], indicating that the external driving electric field can transfer the
ground state to a coherent state. Though the construction for the
state-selective trigger pulse does not use the state-independent excitation
method as mentioned above, this method may be useful when one manipulates
unitarily a Gaussian wave-packet state in time and space by a unitary
propagator of a quadratic Hamiltonian, as can be seen in section 4. This is
because the Hamiltonian (3) is really the specific form of a general
quadratic Hamiltonian. The internal-state-independent excitation method is
also useful in the quantum control process [1]. When the halting-qubit atom
returns back to the left-hand harmonic potential well in the quantum control
process, its motional state needs to be transferred back to the original
ground motional state. Then the state-independent excitation method may help
the quantum control process to realize such a state transfer.

For a neutral atom system both the harmonic potential field and the external
force field may be generated by applying an external driving electric or
magnetic field. It is known that an atom can have an induced electric dipole
moment in the presence of an external electric field [2]. Then a potential
energy $V(x,t)$ of the atom can be generated due to the induced electric
dipole moment in the external electric field $E(x,t)$, which may be
expressed as $V(x,t)=\frac{1}{2}\alpha |E(x,t)|^{2},$ where $\alpha $ is the
atomic polarizability [2]. And hence the interaction $H_{1}(t)$ in Eq. (1)$\ 
$could be obtained from the potential energy $V(x,t).$ Obviously, this
atomic potential energy is internal-state-independent. Thus, the
time-dependent Hamiltonian $H(t)$\ of Eq. (3) for the neutral atom in a
harmonic potential well may also be generated by applying a suitable time-
and space-dependent electric field $E(x,t)$ to the atom. For example, if the
external electric field $E(x,t)\propto ax+b,$ then the potential energy $%
V(x,t)$ contains the linear interaction $H_{1}(t)\propto x.$ On the other
hand, a nuclear spin or a neutral atom which has an intrinsic magnetic
dipole moment may generate a state-dependent force in an inhomogeneous
magnetic field [50]. A nuclear spin or atom in an external magnetic field
can generate the Zeeman effect [2, 50] and the spin energy level is given by 
$E_{m}=-\gamma B(x,t)\hslash m,$ where $B(x,t)$ is the external magnetic
field strength, $m$ the spin magnetic quantum number, and $\gamma $ the
gyromagnetic ratio of the spin. Thus, the interaction between the spin and
the external magnetic field is given by $H_{1}(t)=-\gamma \hslash B(x,t)m.$
When the external magnetic field is not uniform, for example, $%
B(x,t)\varpropto x,$ the spin will be acted on by an external force
generated by the magnetic field $B(x,t)$ if the spin magnetic quantum number 
$m\neq 0.$ Since the quantum number $m$ marks the spin quantum state, the
spin state with $m=0$ is not acted on by the external magnetic field, while
all those spin states with $m\neq 0$ undergo the external magnetic field.
Therefore, the spin interaction $H_{1}(t)$\ may be time- and
state-dependent. Then an inhomogeneous external magnetic field could act as
a state-selective trigger pulse if the halting-qubit atom is chosen as a
nuclear spin (its spin quantum number is even) or a neutral atom with an
intrinsical magnetic dipole moment and the external magnetic field is
designed suitably. \newline
\newline
\newline
{\large 2.2. The off-resonance selective excitation method}

A general method to construct the state-selective trigger pulse is involved
in manipulating and controlling in time and space the center-of-mass motion
and the internal electronic (or spin) motion of an atom as well as the
coupling between the center-of-mass and the internal motion. Since the
coherent-state selective excitation process is dependent upon the specific
internal state of the atom during the state-selective trigger pulse, the
atomic center-of-mass motion must be coupled with the internal electronic
(or spin) motion of the atom in the excitation process. Coherently
manipulating the center-of-mass and the internal motion of the atom as well
as the interaction between the two motions is the fundament for realizing
the atomic laser light cooling and trapping in an atomic ensemble [22],
implementing the quantum computation [23], and preparing and transferring
various quantum states in an atom-ionic system in a harmonic potential field
[24, 25, 26, 27]. The state-selective trigger pulse is also closely related
to the coupling between the atomic center-of-mass and internal motions.
Laser light is a general technique to realize the coupling between the
internal and the center-of-mass motion of an atom. In particular, as one of
the most useful double-photon excitation methods the stimulated Raman
adiabatic passage (STIRAP) method [9] has been extensively used to
coherently manipulate the center-of-mass and the internal motion of an atom
and also used to create and control the coupling between the atomic
center-of-mass and internal motions in an atomic ensemble in quantum
interference experiments [28, 29]. The largest advantage of the STIRAP
method is that the STIRAP method can achieve a complete state-transfer
efficiency in theory and is tolerant to the experimental imperfection and
can avoid the spontaneous emission generated by atomic excited states. For
example, a high internal-state transfer efficiency is achieved
experimentally by the STIRAP method in trapped ions [30]. On the other hand,
the conventional Raman double-photon laser light techniques also have been
used to prepare the ground motional state and various quantum coherent
states and to transfer one atomic internal state to another in an atomic ion
system in a harmonic potential field [18, 27, 31, 32, 33]. The STIRAP method
usually uses a pair of Raman adiabatic laser light beams to couple the
atomic center-of-mass motional state and internal electronic states (or spin
polarization states) so as to realize the interaction between the atomic
center-of-mass and internal motions. In order to generate effectively and
internal-state-selectively the coupling between the two atomic motions the
laser light electromagnetic field should have an oscillatory frequency close
to the resonance frequency of a given pair of the internal electronic (or
spin) states of the atom and far from those resonance frequencies of any
atomic internal-state transitions other than this given resonance frequency.
With this frequency setting the electromagnetic field can induce only the
transition between the given pair of the atomic internal states and does not
induce any other atomic internal-state transitions when the amplitude of the
electromagnetic field is not very large. Thus, this is an
internal-state-selective excitation process. This state-selective excitation
process may be described by the unitary quantum dynamics in theory. Since
this is an internal-state-selective excitation process, one must consider
the atom to be a multi-level physical system. The simplest case is that the
atom is a three-level physical system and two of the three internal energy
levels are irradiated selectively by the external electromagnetic field.
Then in this case the atomic internal Hamiltonian $H(r)$\ may be simply
written as 
\begin{equation}
H(r)=E_{0}|\psi _{0}(r)\rangle \langle \psi _{0}(r)|+E_{1}|\psi
_{1}(r)\rangle \langle \psi _{1}(r)|+E_{2}|\psi _{2}(r)\rangle \langle \psi
_{2}(r)|  \tag{4}
\end{equation}%
where $E_{k}$ and $|\psi _{k}(r)\rangle $ ($k=0,1,2$) are the $k-$th
eigenvalue and eigenstate of the internal Hamiltonian $H(r),$ respectively,
that is, $H(r)|\psi _{k}(r)\rangle =E_{k}|\psi _{k}(r)\rangle .$ For
example, the eigenstates $|\psi _{0}(r)\rangle $ and $|\psi _{1}(r)\rangle $
may be taken as the two hyperfine ground electronic states $|g_{0}\rangle $
and $|g_{1}\rangle $ of an atom, respectively, while $|\psi _{2}(r)\rangle $
is taken as the atomic excited state $|e\rangle .$ Here suppose that the
external electromagnetic field is applied selectively to the two atomic
internal energy levels $|g_{0}\rangle $ and $|e\rangle ,$ while the internal
state $|g_{1}\rangle $ is not affected by the electromagnetic field. \newline
On the other hand, the semiclassical theory of the electromagnetic radiation
and the electric dipole approximation are still suited to describe the
state-selective excitation process in the physical system of the atom plus
the electromagnetic field [2, 34]. In the electric dipole approximation the
interaction between the atom and an externally applied electromagnetic field
can be expressed as $H_{1}(t)=-D\centerdot \mathbf{E}(x,t),$ where $D$ is
the electric dipole operator of the atom, $E(x,t)$ the time-dependent
electric field of the externally applied electromagnetic field, and the
coordinate $x$ the center-of-mass position of the atom. The selective
excitation method may use either the single- or double-frequency (or
double-photon)\ or even multi-photon excitation method. For the single- and
double-frequency selective excitation processes the external electromagnetic
fields are respectively written as 
\begin{equation}
E(x,t)=\frac{1}{2}E_{L0}(t)\exp [i(k_{L0}.x-\omega _{L0}t)]+C.C.  \tag{5}
\end{equation}%
and 
\[
E(x,t)=\frac{1}{2}E_{L0}(t)\exp [i(k_{L0}.x-\omega _{L0}t)] 
\]%
\begin{equation}
+\frac{1}{2}E_{L1}(t)\exp [i(k_{L1}.x-\omega _{L1}t)]+C.C.,  \tag{6}
\end{equation}%
where $E_{Lk}(t),$ $k_{Lk},$ and $\omega _{Lk}$ ($k=0,1$) are the complex
amplitude, wavevector, and oscillatory frequency of the laser light beam,
respectively, and $C.C.$ stands for the complex (or Hermite) conjugate term.
In the single-frequency selective excitation process only one laser light
beam ($E_{L0}(t),k_{L0},\omega _{L0}$) is selectively applied to the two
internal energy levels $|g_{0}\rangle $ and $|e\rangle ,$ while in the
double-frequency selective excitation process a pair of the laser light
beams ($E_{L0}(t),k_{L0},\omega _{L0}$) and ($E_{L1}(t),k_{L1},\omega _{L1}$%
) are selectively applied to the two internal states $|g_{0}\rangle $ and $%
|e\rangle $ simultaneously. Note that the atomic internal state $%
|g_{1}\rangle $ is not affected by any laser light beam in the single- and
double-frequency excitation processes. There is a slight difference between
the double-frequency selective excitation method here and the conventional
STIRAP experiment. In the conventional STIRAP experiment a pair of the Raman
laser light beams usually are applied to two different transitions linking
the two different ground internal states $|g_{0}\rangle $ and $|g_{1}\rangle 
$ to the same excited state $|e\rangle ,$ respectively. Now the electric
dipole interaction $H_{1}(t)$ between the atom and the electric field $%
E(x,t) $ of the external electromagnetic field is given explicitly by [24,
34, 35]%
\[
H_{1}(t)=\hslash (\Omega _{L0}(t)I^{+}+\Omega _{L0}^{\ast }(t)I^{-})\cos
(k_{L0}x-\omega _{L0}t) 
\]%
\[
=\hslash \Omega _{0}(t)\{I^{+}\exp [i(k_{L0}x-\omega _{L0}t-\varphi
_{0}(t))] 
\]%
\begin{equation}
+I^{-}\exp [-i(k_{L0}x-\omega _{L0}t-\varphi _{0}(t))]\}  \tag{7}
\end{equation}%
for the single-frequency selective excitation process, where the second
equality is obtained in the rotating wave approximation, and in the rotating
wave approximation for the double-frequency selective excitation process, 
\[
H_{1}(t)=\hslash \Omega _{0}(t)\{I^{+}\exp [i(k_{0}x-\omega _{0}t-\varphi
_{0}(t))] 
\]%
\[
+I^{-}\exp [-i(k_{0}x-\omega _{0}t-\varphi _{0}(t))]\} 
\]%
\[
+\hslash \Omega _{1}(t)\{I^{+}\exp [i(k_{1}x-\omega _{1}t-\varphi _{1}(t))] 
\]%
\begin{equation}
+I^{-}\exp [-i(k_{1}x-\omega _{1}t-\varphi _{1}(t))]\}  \tag{8}
\end{equation}%
where the two laser light beams may be either counterpropagating ($k_{0}$
and $k_{1}$ have the opposite signs) or copropagating ($k_{0}$ and $k_{1}$
have the same signs), the amplitude $\Omega _{Ll}(t)=\Omega _{l}(t)\exp
[-i\varphi _{l}(t)]$ ($l=0,1$)$,$ and the atomic internal-state operators
are defined by%
\[
2I_{z}=(I^{1}-I^{0}),\text{ }I^{0}=|g_{0}\rangle \langle g_{0}|,\text{ }%
I^{1}=|e\rangle \langle e|,\text{ }I^{+}=|e\rangle \langle g_{0}|,\text{ }%
I^{-}=|g_{0}\rangle \langle e|. 
\]%
These interactions (7)\ and (8) are similar to those in the Jaynes--Cummings
model [35] of an atom plus electromagnetic field system. Now the total
Hamiltonian (1) can be given explicitly if the internal Hamiltonian $H(r)$
(4) and the electric dipole interaction $H_{1}(t)$\ of Eq. (7) or (8) are
inserted into Eq. (1). It should be pointed out that it is possible to apply
an extra laser light beam for each laser light beam above to compensate the
rotating-wave approximation. If each laser light beam in the selective\
excitation process above is replaced with a pair of the laser light beams
with the orthogonal electric field vectors and the suitable phases, one can
eliminate the rotating-wave approximation.\ This means that the Hamiltonian
(8)\ may be constructed exactly.

It is known that the quantum behavior of an atom in the harmonic potential
field and in the presence of the external electromagnetic field may be
described by the complete set of the product basis states $\{|\Psi
_{nk}(x,r)\rangle \}$, 
\begin{equation}
|\Psi _{nk}(x,r)\rangle =|\psi _{n}(x)\rangle |\psi _{k}(r)\rangle ,  \tag{9}
\end{equation}%
where $|\psi _{n}(x)\rangle $ is an eigenstate of the Hamiltonian $H_{0}$
(2) of the harmonic oscillator, which is used to describe the atomic
center-of-mass motion, while $|\psi _{k}(r)\rangle $ is an eigenstate of the
internal Hamiltonian $H(r),$ as can be seen in Eq. (4), which is used to
describe the internal electronic (or spin) motion of the atom. Since the
external electromagnetic field is applied to only the two internal states $%
|g_{0}\rangle $ and $|e\rangle ,$ the time evolution process of the atom in
the harmonic potential well and in the presence of the external
electromagnetic field is described by the unitary propagator 
\begin{equation}
U(t)=T\exp \{-\frac{i}{\hslash }\int_{0}^{t}dt^{\prime }H(t^{\prime
})\}=\exp \{-i\frac{E_{1}t}{\hslash }|g_{1}\rangle \langle g_{1}|\}U_{L}(t) 
\tag{10}
\end{equation}%
where the total Hamiltonian $H(t)$\ is given by Eq. (1) and the unitary
propagator $U_{L}(t)$ is defined as%
\begin{equation}
U_{L}(t)=T\exp \{-\frac{i}{\hslash }\int_{0}^{t}dt^{\prime }H_{L}(t^{\prime
})\}  \tag{10a}
\end{equation}%
with the Hamiltonian $H_{L}(t)$ given by 
\begin{equation}
H_{L}(t)=\frac{p^{2}}{2m}+\frac{1}{2}m\omega
^{2}x^{2}+E_{0}I^{0}+E_{2}I^{1}+H_{1}(t).  \tag{11}
\end{equation}%
Though the Hamiltonian $H_{L}(t)$ is involved in only the two-state subspace
span by the internal states $|g_{0}\rangle $ and $|e\rangle ,$ it\ contains
the term $H_{0}$ (2) and consequently the unitary propagator $U_{L}(t)$
still can affect the atomic product state $|\Psi (x,t)\rangle |g_{1}\rangle
, $ where $|\Psi (x,t)\rangle $ is an atomic center-of-mass motional state.
For example, the unitary propagator $U_{L}(t)$ may have a significant effect
on the product state $|\Psi (x,t)\rangle |g_{1}\rangle $ if the motional
state $|\Psi (x,t)\rangle $ is a superposition state. Actually, the
Hamiltonian $H_{L}(t)$ could not be considered as the Hamiltonian of a
two-level system consisting of the internal states $|g_{0}\rangle $ and $%
|e\rangle $. However, the unitary propagator $U_{L}(t)$ can generate only a
global phase factor for the product state $|\Psi (x,t)\rangle |g_{1}\rangle $
if the motional state $|\Psi (x,t)\rangle $ is an energy eigenstate $|\psi
_{n}(x)\rangle $ of the harmonic oscillator. Now suppose that at the initial
time the halting-qubit atom is in the product state $|\psi _{0}(x)\rangle
|g_{1}\rangle ,$ where $|\psi _{0}(x)\rangle $ is the ground motional state
of the harmonic oscillator. Obviously, the atom is still in the product
state $|\psi _{0}(x)\rangle |g_{1}\rangle $ if it is acted on by the unitary
propagator $U(t)$ of Eq. (10). However, after the atom is transferred to
another internal state $|g_{0}\rangle $ or $|e\rangle $ from the initial
internal state $|g_{1}\rangle ,$ it could be excited to a motional state
that has a higher motional energy than the ground motional state by the
unitary propagator\ $U(t).$ This means in the case that the unitary
propagator $U(t)$ is really state-selective, and the excitation process may
be described directly by the Hamiltonian $H_{L}(t)$ of Eq. (11). Now one
wants to design the external electromagnetic field $E(x,t)$ of Eq. (5)\ or
(6) such that the atomic ground motional state is transferred\ to a standard
coherent state of harmonic oscillator by the unitary propagator $U(t)$ of
Eq. (10) or $U_{L}(t)$ of Eq. (10a). The conventional Raman laser light
beams [18, 19, 31, 32, 33] are often used to generate selectively the
coherent state. They may be understood intuitively below. At the starting
time of the excitation process the atom is in the ground motional state $%
|\psi _{0}(x)\rangle $ and the internal state $|g_{0}\rangle .$ Then the
first Raman laser light beam excites the atom from the internal state $%
|g_{0}\rangle $ to the excited state $|e\rangle $, while the second Raman
laser light beam induces the atom from the excited state $|e\rangle $ to
jump back to the original internal state $|g_{0}\rangle ,$ and at the same
time the initial ground motional state $|\psi _{0}(x)\rangle $ is changed to
a coherent state during the excitation process. Although the atomic internal
state $|g_{0}\rangle $ is not changed after the excitation process, the
atomic motional state is changed from the initial ground state $|\psi
_{0}(x)\rangle $ to the coherent state. Obviously, this coherent state is
indirectly generated through the internal-state transfer pathway $%
|g_{0}\rangle \rightarrow |e\rangle \rightarrow |g_{0}\rangle $ by the two
Raman laser light beams. Therefore, this is an internal-state-dependent
excitation process.

It could be convenient to investigate the internal-state-selective
excitation process based on the Raman laser light beams in the Heisenberg
picture which is often used in the laser spectroscopy [34]. The excitation
process is involved in only the two-state subspace span by the internal
states $|g_{0}\rangle $ and $|e\rangle $ and is governed by the Hamiltonian $%
H_{L}(t)$ of Eq. (11). Then all those operators appearing in the Hamiltonian 
$H_{L}(t)$\ of Eq. (11) are defined as $A(t)=U_{L}(t)^{+}AU_{L}(t)$ in the
Heisenberg picture$,$ where the operator $A$\ may be $I^{0}$, $I^{1}$, $%
I^{+} $, $I^{-}$, $x$, and $p.$ The dynamical equations for these operators
in the Heisenberg picture are given by 
\begin{equation}
i\hslash \frac{d}{dt}A(t)=[A(t),H_{L}(t)].  \tag{12}
\end{equation}%
Below consider a general double-frequency excitation method including those
using the conventional Raman laser light beams. By inserting the Hamiltonian 
$H_{L}(t)$ of Eq. (11)\ with the electric dipole interaction $H_{1}(t)$\ of
Eq. (8) into the Heisenberg equations (12) one obtains [34]\newline
\[
i\hslash \frac{d}{dt}I^{+}(t)=-\hslash \omega _{a}I^{+}(t)+2\hslash \Omega
_{0}(t)I_{z}(t)\exp [-ik_{0}x(t)]\exp [i(\omega _{0}t+\varphi _{0})] 
\]%
\begin{equation}
+2\hslash \Omega _{1}(t)I_{z}(t)\exp [-ik_{1}x(t)]\exp [i(\omega
_{1}t+\varphi _{1})],  \tag{12a}
\end{equation}%
\[
i\hslash \frac{d}{dt}I^{-}(t)=\hslash \omega _{a}I^{-}(t)-2\hslash \Omega
_{0}(t)I_{z}(t)\exp [ik_{0}x(t)]\exp [-i(\omega _{0}t+\varphi _{0})] 
\]%
\begin{equation}
-2\hslash \Omega _{1}(t)I_{z}(t)\exp [ik_{1}x(t)]\exp [-i(\omega
_{1}t+\varphi _{1})],  \tag{12b}
\end{equation}%
\[
i\hslash \frac{d}{dt}I_{z}(t)=\hslash \Omega _{0}(t)\{I^{+}(t)\exp
[i(k_{0}x(t)-\omega _{0}t-\varphi _{0})] 
\]%
\[
-I^{-}(t)\exp [-i(k_{0}x(t)-\omega _{0}t-\varphi _{0})]\} 
\]%
\[
+\hslash \Omega _{1}(t)\{I^{+}(t)\exp [i(k_{1}x(t)-\omega _{1}t-\varphi
_{1})] 
\]%
\begin{equation}
-I^{-}(t)\exp [-i(k_{1}x(t)-\omega _{1}t-\varphi _{1})]\},  \tag{12c}
\end{equation}%
and 
\[
\frac{d}{dt}p(t)=-m\omega ^{2}x(t) 
\]%
\[
-i\hslash k_{0}\Omega _{0}(t)\{I^{+}(t)\exp [i(k_{0}x(t)-\omega
_{0}t-\varphi _{0})] 
\]%
\[
-I^{-}(t)\exp [-i(k_{0}x(t)-\omega _{0}t-\varphi _{0})]\} 
\]%
\[
-i\hslash k_{1}\Omega _{1}(t)\{I^{+}(t)\exp [i(k_{1}x(t)-\omega
_{1}t-\varphi _{1})] 
\]%
\begin{equation}
-I^{-}(t)\exp [-i(k_{1}x(t)-\omega _{1}t-\varphi _{1})]\},  \tag{12d}
\end{equation}%
\begin{equation}
\frac{d}{dt}x(t)=p(t)/m,  \tag{12e}
\end{equation}%
where $E_{0}I^{0}+E_{2}I^{1}=\alpha _{0}E+\hslash \omega _{a}I_{z},$ and $E$
is the $2\times 2$ unit operator, $\alpha _{0}=(E_{2}+E_{0})/2$, $\hslash
\omega _{a}=(E_{2}-E_{0}),$ and $\omega _{a}$ is the resonance frequency of
the two atomic internal energy levels $|g_{0}\rangle $ and $|e\rangle .$ The
former three equations mainly describe the atomic internal motion and the
coupling between the internal and center-of-mass motions of the atom, while
the rest two equations mainly describe the center-of-mass motion and the
coupling of the two motions. The Heisenberg equation set of Eqs. (12)
describes completely the time evolution process of the atom, which is
involved in the atomic internal and center-of-mass motions as well as the
coupling of the two motions. In the absence of the external electromagnetic
field the Heisenberg equations (12) have a simple solution [34]: 
\[
I^{\pm }(t)=I^{\pm }(0)\exp [\pm i\omega _{a}t],\text{ \ }I_{z}(t)=I_{z}(0), 
\]%
\[
a(t)^{+}=a(0)^{+}\exp (i\omega t),\text{ \ }a(t)=a(0)\exp (-i\omega t), 
\]%
where the creation ($a^{+}$) and annihilation ($a$) operators defined
through 
\begin{equation}
x(t)=\sqrt{\frac{\hslash }{2m\omega }}(a(t)^{+}+a(t)),\text{ }p(t)=i\sqrt{%
\frac{1}{2}\hslash \omega m}(a(t)^{+}-a(t)).  \tag{13}
\end{equation}%
This simple solution is called the uncoupling solution to the Heisenberg
equations (12).

Now the off-resonance excitation method is introduced below. The
off-resonance excitation means that the two atomic internal states $%
|g_{0}\rangle $ and $|e\rangle $ are irradiated by weak and off-resonance
laser light beams. This also means that for the off-resonance excitation
using the Raman adiabatic laser light beams the Rabi frequencies $\Omega
_{0}(t)$ and $\Omega _{1}(t)$ of the two Raman adiabatic laser light beams
are slowly varying and much less than the frequency offsets $(\omega
_{a}-\omega _{0})$ and $(\omega _{a}-\omega _{1})$. Obviously, in the
off-resonance excitation the solution to the Heisenberg equations (12)
should be close to the uncoupling solution and hence the operators $\{I^{\pm
}(t)\exp [\mp i\omega _{a}t]\}$ should be close to the operator $I^{\pm
}(0). $ Therefore, the time derivatives of the operators $\{I^{\pm }(t)\exp
[\mp i\omega _{a}t]\}$ are close to zero. Now one may make a unitary
transformation [33]:%
\begin{equation}
\hat{I}^{\pm }(t)=\exp (-i\omega _{a}I_{z}(t)t)I^{\pm }(t)\exp (i\omega
_{a}I_{z}(t)t)=I^{\pm }(t)\exp [\mp i\omega _{a}t].  \tag{14}
\end{equation}%
Then the first two Heisenberg equations (12a) and (12b) are reduced
respectively to the forms%
\[
\frac{d}{dt}\hat{I}^{+}(t)=-2i\Omega _{0}(t)I_{z}(t)\exp [-ik_{0}x(t)]\exp
[-i(\omega _{a}-\omega _{0})t+i\varphi _{0}] 
\]%
\begin{equation}
-2i\Omega _{1}(t)I_{z}(t)\exp [-ik_{1}x(t)]\exp [-i(\omega _{a}-\omega
_{1})t+i\varphi _{1})]  \tag{15a}
\end{equation}%
and 
\[
\frac{d}{dt}\hat{I}^{-}(t)=2i\Omega _{0}(t)I_{z}(t)\exp [ik_{0}x(t)]\exp
[i(\omega _{a}-\omega _{0})t-i\varphi _{0}] 
\]%
\begin{equation}
+2i\Omega _{1}(t)I_{z}(t)\exp [ik_{1}x(t)]\exp [i(\omega _{a}-\omega
_{1})t-i\varphi _{1}].  \tag{15b}
\end{equation}%
Obviously, the time derivatives of the operators $\{\hat{I}^{\pm }(t)\}$ are
close to zero when the Rabi frequencies $\Omega _{0}(t)$ and $\Omega _{1}(t)$
are close to zero. If the frequency offsets $\{|(\omega _{a}-\omega _{l})|\}$
are much greater than the Rabi frequencies $\{\Omega _{l}(t)\}$ ($l=0,1$),
the atomic motional velocity $p(t)/m$ is small in the harmonic potential
well, and the time varying of the Rabi frequencies is slow (the Raman
adiabatic laser light beams satisfy the condition), then the operators $\{%
\hat{I}^{\pm }(t)\}$ may be obtained approximately by integrating the two
equations (15), respectively, because the oscillatory terms $\exp [\pm
i(\omega _{l}-\omega _{a})t]$ ($l=0,1$) will generate a dominating
contribution to the two integrals. Now by integrating first the two
equations (15), then making integration by part, and then by using Eq. (14)
one obtains the two operators$:$ 
\[
I^{+}(t)-\exp (i\omega _{a}t)I^{+}(0)=\frac{2\exp (i\varphi _{0})}{(\omega
_{a}-\omega _{0})}\Omega _{0}(t)I_{z}(t)\exp [-ik_{0}x(t)]\exp [i\omega
_{0}t] 
\]%
\begin{equation}
+\frac{2\exp (i\varphi _{1})}{(\omega _{a}-\omega _{1})}\Omega
_{1}(t)I_{z}(t)\exp [-ik_{1}x(t)]\exp [i\omega _{1}t]+E(I^{+}),  \tag{16a}
\end{equation}%
\[
I^{-}(t)-\exp (-i\omega _{a}t)I^{-}(0)=\frac{2\exp (-i\varphi _{0})}{(\omega
_{a}-\omega _{0})}\Omega _{0}(t)I_{z}(t)\exp [ik_{0}x(t)]\exp [-i\omega
_{0}t] 
\]%
\begin{equation}
+\frac{2\exp (-i\varphi _{1})}{(\omega _{a}-\omega _{1})}\Omega
_{1}(t)I_{z}(t)\exp [ik_{1}x(t)]\exp [-i\omega _{1}t]+E(I^{-})  \tag{16b}
\end{equation}%
For simplicity, here suppose that the phases $\varphi _{0}$ and $\varphi
_{1} $ are time-independent and at the initial time both the amplitudes of
the two Raman adiabatic laser light beams are zero, that is, $\Omega
_{0}(0)=\Omega _{1}(0)=0.$ Since these terms $\{\exp (\pm i\omega
_{a}t)I^{\pm }(0)\}$ are of the uncoupling solution, the right-hand sides of
the equations (16) measure the deviation of the real solution from the
uncoupling solution. The terms $E(I^{+})$ and $E(I^{-})$ in Eq. (16) are
error terms which are higher-order terms of the frequency offsets ($(\omega
_{a}-\omega _{l})^{-2}$ $(l=0,1),$ ($\omega _{a}-\omega _{l})^{-1}(\omega
_{a}-\omega _{l})^{-1},$ etc.). The error terms $E(I^{+})$ and $E(I^{-})$
can be neglected in the first-order approximation. That is, only the
first-order terms which are proportional to the inverse frequency offsets ($%
|\omega _{a}-\omega _{l}|^{-1},$ $l=0,1$) are retained in the operators $%
\{I^{\pm }(t)\}$ in the first-order approximation. Now inserting the
operators $\{I^{\pm }(t)\}$ (16a) and (16b)\ of the first-order
approximation into the Heisenberg equation (12c) one obtains, up to the
first-order approximation, 
\[
I_{z}(t)-I_{z}(0)=-\frac{\Omega _{0}(t)\exp [-i\varphi _{0}]}{(\omega
_{a}-\omega _{0})}I^{+}(0)\exp [ik_{0}x(t)]\newline
\exp [i(\omega _{a}-\omega _{0})t] 
\]%
\begin{equation}
-\frac{\Omega _{1}(t)\exp [-i\varphi _{1}]}{(\omega _{a}-\omega _{1})}%
I^{+}(0)\exp [ik_{1}x(t)]\newline
\exp [i(\omega _{a}-\omega _{1})t]+C.C.  \tag{16c}
\end{equation}%
where $C.C.$ stands for the hermite conjugate of the first two terms on the
right-hand side of Eq. (16c). The two Heisenberg equations (12d) and (12e)
then can be solved with the help of the first-order approximation operators $%
\{I^{\pm }(t)\}$ (16a) and (16b) and $I_{z}(t)$ (16c) and one obtains, in
the first-order approximation,%
\[
a(t)^{+}-\exp (i\omega t)a(0)^{+}=iI_{z}(0)\sqrt{\frac{2\hslash }{m\omega }}%
\int_{0}^{t}dt^{\prime }\{\Omega _{eff}(t^{\prime }) 
\]%
\begin{equation}
\times \exp [-i\omega (t^{\prime }-t)]\sin [\Delta \omega t^{\prime }+\Delta
\varphi ]\},  \tag{16d}
\end{equation}%
\[
a(t)-\exp (-i\omega t)a(0)=-iI_{z}(0)\sqrt{\frac{2\hslash }{m\omega }}%
\int_{0}^{t}dt^{\prime }\{\Omega _{eff}(t^{\prime }) 
\]%
\begin{equation}
\times \exp [i\omega (t^{\prime }-t)]\sin [\Delta \omega t^{\prime }+\Delta
\varphi ]\},  \tag{16e}
\end{equation}%
where the Lamb-Dicke limit $||\Delta kx(t)||<<1$ and the condition $m\omega
^{2}/\hslash >>|\Delta k\Omega _{eff}(t)|$ have been used; $\Delta
k=k_{0}-k_{1},$ $\Delta \omega =\omega _{0}-\omega _{1},$ $\Delta \varphi
=\varphi _{0}-\varphi _{1};$ and the parameter $\Omega _{eff}(t)$ is given by%
\[
\Omega _{eff}(t)=\frac{2k_{0}\Omega _{0}(t)\Omega _{1}(t)}{(\omega
_{a}-\omega _{1})}-\frac{2k_{1}\Omega _{0}(t)\Omega _{1}(t)}{(\omega
_{a}-\omega _{0})}. 
\]%
The condition $m\omega ^{2}>>|\hslash \Delta k\Omega _{eff}(t)|$ means that
the effect of the Raman laser light beams on the harmonic potential field $%
(m\omega ^{2}x^{2}/2)$ can be neglected. Now by inserting the first-order
approximation operators $\{I^{\pm }(t)\}$ of Eq. (16a) and (16b) into the
electric dipole interaction $H_{1}(t)$ (8) and then inserting Eq. (8)\ into
Eq. (11)\ the Hamiltonian $H_{L}(t)$ of Eq. (11) can be written as,\ in the
first-order approximation, 
\[
H_{L}(t)=\alpha _{0}E+\hslash \omega \lbrack a^{+}(t)a(t)+\frac{1}{2}%
]+\hslash \Omega _{a}(t)I_{z}(t) 
\]%
\[
+2\hslash \Delta k[\Omega _{e0}(t)+\Omega _{e1}(t)]I_{z}(t)x(t)\sin [\Delta
\omega t+\Delta \varphi ] 
\]%
\[
+\hslash \Omega _{0}(t)\{\exp (-i\varphi _{0})I^{+}(0)\exp [ik_{0}x(t)]\exp
[i(\omega _{a}-\omega _{0})t] 
\]%
\[
+\exp (i\varphi _{0})\exp [-ik_{0}x(t)]I^{-}(0)\exp [-i(\omega _{a}-\omega
_{0})t]\} 
\]%
\[
+\hslash \Omega _{1}(t)\{\exp (-i\varphi _{1})I^{+}(0)\exp [ik_{1}x(t)]\exp
[i(\omega _{a}-\omega _{1})t] 
\]%
\begin{equation}
+\exp (i\varphi _{1})\exp [-ik_{1}x(t)]I^{-}(0)\exp [-i(\omega _{a}-\omega
_{1})t]\}  \tag{17}
\end{equation}%
where the expansion $\exp [\pm i\Delta kx(t)]=1\pm i\Delta kx(t)+...$ in the
Lamb-Dicke limit has been used, the parameter $\Omega _{a}(t)$ is given by 
\[
\Omega _{a}(t)=\omega _{a}+\frac{4\Omega _{0}(t)^{2}}{(\omega _{a}-\omega
_{0})}+\frac{4\Omega _{1}(t)^{2}}{(\omega _{a}-\omega _{1})}+2[\Omega
_{e0}(t)+\Omega _{e1}(t)]\cos (\Delta \omega t+\Delta \varphi ), 
\]%
and the effective Rabi frequency $\Omega _{el}(t)=2\Omega _{0}(t)\Omega
_{1}(t)/(\omega _{a}-\omega _{l})$ ($l=0,1$) which was also obtained in
Refs. [36]. It is known that the Lamb-Dicke parameter $\eta $ is defined
through $kx(t)=\eta (a(t)^{+}+a(t))$. The Lamb-Dicke parameter $\eta _{l}=%
\sqrt{(\hbar ^{2}k_{l}^{2})/(2m\hslash \omega )}$ with the wave vector value 
$k_{l}=\omega _{l}/c$ ($l=0,1$) may be large if the frequencies $\omega _{0}$
and $\omega _{1}$ of the two Raman laser light beams are near the transition
frequency $\omega _{a}$ $(\thicksim 10^{15})$ between the atomic internal
states $|g_{0}\rangle $ and $|e\rangle .$ However, the wave vector
difference $|\Delta k|=|k_{0}-k_{1}|$ may be very small and its value can be
controlled in experiment. For example, $|\Delta k|=|\Delta \omega
|/c\thicksim \omega /c<<\omega _{a}/c,$ since the oscillatory frequency of
the harmonic oscillator $\omega \sim 10^{8}$ is much less than the atomic
internal-state transition frequency $\omega _{a}\thicksim 10^{15}$. In this
case the Lamb-Dicke parameter may be very small: $\eta =\sqrt{\hbar
^{2}(\Delta k)^{2}/(2m\hslash \omega )}<<1,$ and hence $||\Delta kx(t)||<<1$.

It seems that one may directly obtain the first-order effective Hamiltonian
from Eq. (17)\ by omitting those terms of the operators $\{I^{\pm }(0)\}$ on
the right-hand side of Eq. (17), but the correct treatment is that one
should use the first-order approximation solution of the Heisenberg
equations (12) which contains the operators $I_{z}(t),$ $a(t)^{+},$ and $%
a(t) $ of the equations (16c), (16d), and (16e) to reduce the Hamiltonian
(17) to the first-order effective Hamiltonian. The Hamiltonian (17) may be
divided into two parts, the first part does not contain the operators $%
\{I^{\pm }(0)\},$ while the second contains. Note that the effective Rabi
frequency $\Omega _{a}(t)$ in the Hamiltonian (17) is close to $\omega _{a}$
when the Rabi frequency $\Omega _{l}(t)$ ($l=0,1$) is much smaller than $%
|(\omega _{a}-\omega _{l})|.$ Then the part that contains the operators $%
\{I^{\pm }(0)\}$ is nonsecular, since its components generally contain
oscillatory phase factors $\exp [\pm i(\omega _{a}-\omega _{l})t]$ ($l=0,1$)
instead of $\exp [\pm i\omega _{l}t]$ (see Eq. (17)) and in the rotating
frame defined by the Hamiltonian $\omega _{l}I_{z}$ or $\omega _{a}I_{z}$
they still contain largely oscillatory phase factors $\exp [\pm i\omega
_{a}t]$, $\exp [\pm i\omega _{l}t],$ etc., instead of the slowly varying
phase factors $\exp [\pm i\Delta \omega t]$. Thus, the contribution of the
nonsecular part to the off-resonance excitation is usually small. If one
chooses the suitable parameters of the Raman laser light beams such that the
absolute amplitudes in front of the oscillatory phase factors $\exp [\pm
i(\omega _{a}-\omega _{l})t]$ ($l=0,1$) in the nonsecular part are much
smaller than $|(\omega _{a}-\omega _{l})|$, then the contribution of the
nonsecular part to the off-resonance excitation may be neglected. For
convenience to discuss below, here the parameters of the Raman adiabatic
laser light beams are chosen suitably such that the first part is much more
important than the second and hence the nonsecular part that contains the
operators $\{I^{\pm }(0)\}$ may be neglected. Then in this case the
Hamiltonian (17)\ is reduced approximately to the simple form 
\begin{equation}
H_{L}(t)=\hslash \omega (a^{+}a+\frac{1}{2})+\hslash \Omega
_{a}(t)I_{z}+\hslash \lbrack \beta (t)a^{+}+\beta (t)^{\ast }a]I_{z} 
\tag{18}
\end{equation}%
where the unity operator term is omitted and the parameter $\beta (t)$ is
given by 
\[
\beta (t)=-i\sqrt{\frac{2\hslash \omega }{m}}\int_{0}^{t}dt^{\prime }\Omega
_{eff}(t^{\prime })\sin [\Delta \omega t^{\prime }+\Delta \varphi ]\exp
(i\omega t^{\prime }) 
\]%
\[
+\Delta k\sqrt{\frac{2\hslash }{m\omega }}[\Omega _{e0}(t)+\Omega
_{e1}(t)]\sin [\Delta \omega t+\Delta \varphi ]\exp (i\omega t). 
\]%
In the parameter $\beta (t)$ the integral containing the parameter $\Omega
_{eff}(t)$ usually is not considered [36]. If the two frequencies offsets $%
\{\omega _{a}-\omega _{l}\}$ ($l=0,1$) have the opposite sign to each other
such that the parameter $|\Omega _{eff}(t)|$ is large, then the dominating
contribution to the parameter $\beta (t)$ may come from the integral
containing the parameter $\Omega _{eff}(t).$ The first-order effective
Hamiltonian (18) can be used to excite a coherent state of harmonic
oscillator. The cross operator $[\beta (t)a^{+}+\beta (t)^{\ast }a]I_{z}$ in
the Hamiltonian $H_{L}(t)$ (18)\ is responsible for the
internal-state-selective excitation of a coherent state. This will be
interpreted below. In the interaction representation defined by the
harmonic\ oscillator Hamiltonian $H_{0}=\hslash \omega (a^{+}a+1/2)$ the
Hamiltonian (18)\ is transformed to [11]%
\[
H_{LI}(t)=\hslash \Omega _{a}(t)I_{z}+i\hslash \lbrack z(t)a^{+}-z(t)^{\ast
}a]I_{z} 
\]%
where the complex number $z(t)=\beta (t)\exp [i\omega t-i\pi /2].$ The
propagator of the Hamiltonian $H_{LI}(t)$ in the interaction representation
is 
\[
U_{LI}(t)=\exp [-iI_{z}\int_{0}^{t}dt^{\prime }\Omega _{a}(t^{\prime
})]D(\mu (t)) 
\]%
where the unitary displacement operator $D(\mu (t))$ is defined as [11]%
\[
D(\mu (t))=\exp [\mu (t)a^{+}-\mu (t)^{\ast }a]=T\exp
\{I_{z}\int_{0}^{t}dt^{\prime }[z(t^{\prime })a^{+}-z(t^{\prime })^{\ast
}a]\}. 
\]%
Note that here $\mu (t)$ of the operator $D(\mu (t))$ is really an operator
and may be expressed $\mu (t)=\mu _{1}(t)E+2I_{z}\mu _{2}(t)$ [51]$,$ where $%
\mu _{l}(t)$ ($l=1,2)$ is a complex number, $E=(|e\rangle \langle
e|+|g_{0}\rangle \langle g_{0}|),$ and $I_{z}=(|e\rangle \langle
e|-|g_{0}\rangle \langle g_{0}|)/2.$ It is known that the unitary
displacement operator $D(z(t))$ ($z(t)$ is a complex number) can excite a
standard coherent state from the ground state of a harmonic oscillator [11].
Since the operator $\mu (t)$ acts on only the two atomic internal states $%
|e\rangle $ and $|g_{0}\rangle ,$ the unitary displacement operator $D(\mu
(t))$ can excite the coherent state of the atom in the harmonic potential
well only when the atom is in the internal-state subspace span by the two
internal states $|e\rangle $ and $|g_{0}\rangle .$ This indicates that the
unitary propagator $U_{LI}(t)$ can excite the coherent state in an
internal-state dependent form and hence is internal-state-selective. In
order to see more clearly how the first-order effective Hamiltonian (18) can
be used to excite selectively the coherent state one may assume that the
parameter $\beta (t)$ is real, which could be achieved by choosing suitably
the parameters of the Raman laser light beams. Then the Hamiltonian (18)\ is
written as, with the help of Eqs. (13), 
\begin{equation}
H_{L}(t)=\frac{p^{2}}{2m}+\frac{1}{2}m\omega ^{2}x^{2}+\hslash \Omega
_{a}(t)I_{z}+f(t)I_{z}x  \tag{19}
\end{equation}%
where the force function $f(t)=\sqrt{2m\hslash \omega }\beta (t).$ The last
term on the right-hand side of Eq. (19), which is proportional to the
product operator $I_{z}x,$ is responsible for the internal-state-selective
excitation for the coherent state. In fact, the cross term, for example, the
product operator $I_{\mu }x$ ($\mu =x,y,z$) (there are also other cross
terms $I_{\mu }y$ and $I_{\mu }z$ in three dimensions$),$ reflects the
coupling between the atomic center-of-mass and internal motions. In order to
excite efficiently the coherent state the varying frequency of the force
function $f(t)$\ should be close to the oscillatory frequency $\omega $ of
the harmonic oscillator. Now\ one may write the unitary propagator $U_{L}(t)$
for the first-order effective Hamiltonian $H_{L}(t)$ of Eq. (19) as the form%
\begin{equation}
U_{L}(t)=\exp \{-iI_{z}\int_{0}^{t}dt^{\prime }\Omega _{a}(t^{\prime
})\}T\exp \{-\frac{i}{\hslash }\int_{0}^{t}dt^{\prime }H_{f}(t^{\prime })\} 
\tag{20}
\end{equation}%
where the first-order effective Hamiltonian $H_{f}(t)$ is defined by%
\begin{equation}
H_{f}(t)=\frac{p^{2}}{2m}+\frac{1}{2}m\omega ^{2}x^{2}+f(t)I_{z}x  \tag{21}
\end{equation}%
It is clear from Eq. (3)\ that the first-order effective Hamiltonian $%
H_{f}(t)$ of Eq. (21) is very similar to the Hamiltonian (3)\ of a forced
harmonic oscillator. The Hamiltonian $H_{f}(t)$ is responsible for the
state-selective excitation of a coherent state, as can be seen below.

It is known that at the initial time of the state-selective trigger pulse
[1] the halting-qubit atom may be either in the product state $|\psi
_{0}(x)\rangle |g_{1}\rangle $ (the atom is in the internal state $%
|g_{1}\rangle $ and the ground motional state $|\psi _{0}(x)\rangle $) or in
the product state $|\psi _{0}(x)\rangle |g_{0}\rangle $ (or $|\psi
_{0}(x)\rangle |e\rangle $), while the two laser light beams are applied
selectively to the two internal states $|g_{0}\rangle $ and $|e\rangle $ and
do not have any effect on other atomic internal states including the state $%
|g_{1}\rangle .$ Consider the first case that at the initial time the
halting-qubit atom is in the product state $|\psi _{0}(x)\rangle
|g_{1}\rangle .$ The time evolution process for the atom is expressed as 
\[
|\Psi (x,r,t)\rangle =U(t)|\psi _{0}(x)\rangle |g_{1}\rangle =\exp
\{-iE_{1}t/\hslash \}U_{L}(t)|\psi _{0}(x)\rangle |g_{1}\rangle . 
\]%
Since the atomic operator $I_{z}=(|e\rangle \langle e|-|g_{0}\rangle \langle
g_{0}|)/2,$ one has $I_{z}|g_{1}\rangle =0.$ Then it follows from Eq. (19)$\ 
$that $H_{L}(t)|\psi _{0}(x)\rangle |g_{1}\rangle =H_{0}|\psi _{0}(x)\rangle
|g_{1}\rangle .$ Note that $H_{0}|\psi _{0}(x)\rangle =\frac{1}{2}\hslash
\omega |\psi _{0}(x)\rangle .$ Thus, the time evolution process is reduced
to the form%
\[
|\Psi (x,r,t)\rangle =\exp [-iE_{1}t/\hslash ]\exp [-i\omega t/2]|\psi
_{0}(x)\rangle |g_{1}\rangle . 
\]%
Up to a global phase factor the state $|\Psi (x,r,t)\rangle $ is really the
initial product state $|\psi _{0}(x)\rangle |g_{1}\rangle .$ This shows that
the Raman laser light beams do not have any significant effect on the
initial product state $|\psi _{0}(x)\rangle |g_{1}\rangle .$ On the other
hand, for the second case that at the initial time the halting-qubit atom is
in the product state $|\psi _{0}(x)\rangle |g_{0}\rangle $ the time
evolution process for the atom is expressed as 
\[
|\Psi (x,r,t)\rangle =\exp \{-iI_{z}\int_{0}^{t}dt^{\prime }\Omega
_{a}(t^{\prime })\}T\exp \{-\frac{i}{\hslash }\int_{0}^{t}dt^{\prime
}H_{f}(t^{\prime })\}|\psi _{0}(x)\rangle |g_{0}\rangle . 
\]%
Since $I_{z}|g_{0}\rangle =-(1/2)|g_{0}\rangle ,$ the wave function is
reduced to the form 
\[
|\Psi (x,r,t)\rangle =\exp \{i\frac{1}{2}\int_{0}^{t}dt^{\prime }\Omega
_{z}(t^{\prime })\}T\exp \{-\frac{i}{\hslash }\int_{0}^{t}dt^{\prime
}H_{f}^{-}(t^{\prime })\}|\psi _{0}(x)\rangle |g_{0}\rangle 
\]%
where the Hamiltonian $H_{f}^{-}(t)$ is given by 
\[
H_{f}^{-}(t)=\frac{p^{2}}{2m}+\frac{1}{2}m\omega ^{2}x^{2}-\frac{1}{2}f(t)x. 
\]%
One sees that the Hamiltonian $H_{f}^{-}(t)$ is really the Hamiltonian (3)
of a forced harmonic oscillator. Thus, the ground state $|\psi
_{0}(x)\rangle $ will be transferred to a coherent state under the
Hamiltonian $H_{f}^{-}(t)$ [13, 17]. In an analogous way, one can prove that
when the halting-qubit atom is in the product state $|\psi _{0}(x)\rangle
|e\rangle $ at the initial time, the time evolution process is given by 
\[
|\Psi (x,r,t)\rangle =\exp \{-i\frac{1}{2}\int_{0}^{t}dt^{\prime }\Omega
_{z}(t^{\prime })\}T\exp \{-\frac{i}{\hslash }\int_{0}^{t}dt^{\prime
}H_{f}^{+}(t^{\prime })\}|\psi _{0}(x)\rangle |e\rangle 
\]%
where $I_{z}|e\rangle =(1/2)|e\rangle $ has been used and the Hamiltonian $%
H_{f}^{+}(t)$ is given by 
\[
H_{f}^{+}(t)=\frac{p^{2}}{2m}+\frac{1}{2}m\omega ^{2}x^{2}+\frac{1}{2}f(t)x. 
\]%
The Hamiltonian $H_{f}^{+}(t)$ is also the Hamiltonian (3) of a forced
harmonic oscillator and hence the initial ground state $|\psi _{0}(x)\rangle 
$ will be excited to a coherent state under the Hamiltonian $H_{f}^{+}(t)$
[17]. Here gives a summary for the above discussion for the off-resonance
excitation process. The product state of the atom will keep unchanged under
the Raman laser light beams when the atom is in the product state $|\psi
_{0}(x)\rangle |g_{1}\rangle $ at the initial time. When the atom is changed
to the internal state $|g_{0}\rangle $ or $|e\rangle $ from the original
internal state $|g_{1}\rangle $ and hence in the product state $|\psi
_{0}(x)\rangle |g_{0}\rangle $ or $|\psi _{0}(x)\rangle |e\rangle $ at the
starting time of the excitation process, the atom will be excited to a
coherent state by the Raman laser light beams and moreover the atomic
internal state keeps unchanged during the excitation process. Therefore,
this coherent-state excitation process is clearly a state-selective
excitation process. Such a pulse sequence consisting of the two Raman laser
light beams could act as the state-selective trigger pulse to realize the
reversible and unitary halting protocol.

The off-resonance selective excitation using the conventional Raman
adiabatic laser light beams is a simple technique to prepare approximately a
coherent state of an atom or atomic ion in the harmonic potential field. The
above theoretic investigation for the off-resonance selective excitation
shows ones clearly the mechanism for a Raman adiabatic pulse sequence to
excite selectively the coherent state of an atom in the harmonic potential
field. The similar theoretical analysis also can be seen in a number of
references [19, 20, 25a, 31, 33, 45]. The state-selective excitation process
is indeed involved in the coupling between the atomic center-of-mass motion
and the internal electronic (or spin) motion of the atom, since there is the
cross term $xI_{z}$ in the effective Hamiltonian $H_{L}(t)$ of Eq. (19)
which is responsible for the state-selective excitation of the coherent
state. The coupling between the atomic center-of-mass and internal motions
is also a general mechanism for the atomic laser cooling [22] and the
decelerating and accelerating processes based on the STIRAP method of a free
atom [8]. However, if one compares the present theoretic treatment for an
atom in the harmonic potential well to that one for a free atom [8], one can
see their difference is quite large. The reason for this is that a motional
state of the atom in the harmonic potential field is discrete instead of
continuous. Though the ground motional state of the atom in the harmonic
potential well has a Gaussian shape, the excited motional state which is
generated from the ground motional state by the Raman pulse sequence could
not have a Gaussian shape and tends to have a wave-packet shape quite
different from a Gaussian shape. For the off-resonance selective excitation
above the excited motional state has a Gaussian shape in the first-order
approximation, but in a general case it does not have a Gaussian shape and
its shape could be complicated. Thus, in this sense the off-resonance
selective excitation mentioned above generally is not an optimal technique
to construct the state-selective trigger pulse. But it could be considered
as a starting point to develop further a better technique to generate
state-selectively the standard coherent state of harmonic oscillator. One
could exploit the average Hamiltonian theory [37, 39, 54] or the numerical
optimization method [38] based on the Heisenberg equations (12) to improve
the off-resonance excitation technique. The state-selective excitation pulse
obtained with these optimal methods could be more useful in practice. Of
course, it must be pointed out that a state-selective excitation pulse is
not always equal to a state-selective trigger pulse. \newline
\newline
\newline
{\large 2.3. The coherent average method}

Here suggests the programming operator composition method to construct the
state-selective trigger pulse in theory. This method, which has been used
extensively to construct NMR multi-pulse sequences, is also called the
coherent average method in nuclear magnetic resonance spectroscopy [54]. The
theoretical basis of the method is the famous Baker-Campbell-Hausdorff (BCH)
formula [39, 41] and the Trotter-Suzuki formalism [40, 42]. The BCH formula
is also the theoretical basis of the average Hamiltonian theory [37, 54]. It
is necessary to show that the state-selective trigger pulse can be
constructed theoretically in an error as small as pleased so as to show that
both the reversible and unitary halting protocol that is insensitive to its
input state and the efficient quantum search process are feasible. Here
consider the double-frequency selective excitation method, which may be
different from the off-resonance selective excitation method based on the
Raman adiabatic laser light beams in the previous subsection. The method
uses simply two conventional amplitude- and phase-modulation laser light
beams (the plane wave electromagnetic fields), which may not be Raman
adiabatic laser light beams, to excite selectively the atomic internal
states so as to set up the coupling between the center-of-mass and the
internal motion of the atom. In order to treat conveniently the
double-frequency state-selective excitation process in the three-level atom
system here consider only the special case that the two laser light beams
have the specific parameter settings as given below. It is known the
electric dipole interaction between the atom and a pair of laser light beams
is given by Eq. (8). Now suppose that the two laser light beams are
amplitude- and phase-modulating such that their Rabi frequencies and phases
satisfy the following match condition: 
\begin{equation}
\Omega _{0}(t)=\Omega _{1}(t),\text{ }\varphi _{0}(t)=\alpha +\gamma ,\text{ 
}\varphi _{1}(t)=(\omega _{0}-\omega _{1})t-\alpha +\gamma .  \tag{22}
\end{equation}%
Then by using the match condition (22)\ the electric dipole interaction of
Eq. (8) is reduced to the form \newline
\[
H_{1}(t)=2\hslash \Omega _{0}(t)\exp (-i\omega _{0}t)I^{+}\exp [i\frac{1}{2}%
(k_{0}+k_{1})x-i\gamma ]\cos [\frac{1}{2}(k_{0}-k_{1})x-\alpha ] 
\]%
\begin{equation}
+2\hslash \Omega _{0}(t)\exp (i\omega _{0}t)I^{-}\exp [-i\frac{1}{2}%
(k_{0}+k_{1})x+i\gamma ]\cos [\frac{1}{2}(k_{0}-k_{1})x-\alpha ]  \tag{23}
\end{equation}%
where the phases $\alpha $ and $\gamma $ can be set suitably in experiment.
The theoretical treatment could become more convenient for the
state-selective excitation process when the electric dipole interaction $%
H_{1}(t)$\ takes the form of Eq. (23), since the electric dipole interaction
(23) is modulated by a single frequency $\omega _{0}.$ On the other hand, it
could be convenient to treat the state-selective excitation process in the
rotating reference frame. It is known that the Schr\"{o}dinger equation for
a quantum system in the interaction representation can be written as [2]%
\[
i\hslash \frac{\partial }{\partial t}\Psi _{I}(x,t)=H_{I}(t)\Psi _{I}(x,t) 
\]%
where the state $\Psi _{I}(x,t)$ of the quantum system in the interaction
representation is related to the state $\Psi _{L}(x,t)$ in the laboratory
frame by the unitary transformation: $\Psi _{I}(x,t)=U_{0}(t)^{+}\Psi
_{L}(x,t),$ and the Hamiltonian $H_{I}(t)$ of the quantum system in the
interaction representation to the Hamiltonian $H(t)=H_{0}(t)+H_{1}(t)$ in
the laboratory frame by 
\begin{equation}
H_{I}(t)=U_{0}(t)^{+}H_{1}(t)U_{0}(t)  \tag{24}
\end{equation}%
where the unitary propagator $U_{0}(t)$ is defined as 
\begin{equation}
U_{0}(t)=T\exp \{-\frac{i}{\hslash }\int_{0}^{t}dt^{\prime }H_{0}(t^{\prime
})\}.  \tag{25}
\end{equation}%
The total unitary propagator of the quantum system then is written as%
\begin{equation}
U(t)=T\exp \{-\frac{i}{\hslash }\int_{0}^{t}dt^{\prime }[H_{0}(t^{\prime
})+H_{1}(t^{\prime })]\}=U_{0}(t)U_{I}(t)  \tag{26}
\end{equation}%
where the unitary propagator of the quantum system in the interaction
representation is defined as%
\begin{equation}
U_{I}(t)=T\exp \{-\frac{i}{\hslash }\int_{0}^{t}dt^{\prime }H_{I}(t^{\prime
})\}.  \tag{27}
\end{equation}%
One may choose a suitable rotating reference frame or the interaction
representation for convenient treatment of the state-selective excitation
process. First of all, the atomic internal Hamiltonian of Eq. (4)\ is
rewritten as $H(r)=\alpha _{0}E+\hslash \omega _{a}I_{z}+(E_{1}-\alpha
_{0})|g_{1}\rangle \langle g_{1}|,$ where $E$ is the $3\times 3$ unity
matrix. Then by neglecting the unity operator the total Hamiltonian (1)\ for
the three-state atom system is rewritten as 
\begin{equation}
H(t)=H_{0}+\hslash \omega _{a}I_{z}+(E_{1}-\alpha _{0})(|g_{1}\rangle
\langle g_{1}|)+H_{1}(t)  \tag{28}
\end{equation}%
where $H_{0}$ and $H_{1}(t)$ are given by Eqs. (2) and (23), respectively.
Now the rotating reference frame may be defined by the atomic internal
Hamiltonian $H_{0}(r)=\hslash \omega _{0}I_{z}+(E_{1}-\alpha
_{0})(|g_{1}\rangle \langle g_{1}|).$ In the rotating frame the wave
function is 
\[
\Psi _{I}(x,t)=\exp [i(E_{1}-\alpha _{0})(|g_{1}\rangle \langle
g_{1}|)t/\hslash ]\exp (i\omega _{0}I_{z}t)\Psi _{L}(x,t), 
\]%
and the total Hamiltonian $H(t)$ of Eq. (28)\ of the atom system is replaced
with the Hamiltonian $H_{I}(t)$: \newline
\[
H_{I}(t)=\frac{p^{2}}{2m}+\frac{1}{2}m\omega ^{2}x^{2}+\hslash (\omega
_{a}-\omega _{0})I_{z} 
\]%
\[
+2\hslash \Omega _{0}(t)I^{+}\exp [i\frac{1}{2}(k_{0}+k_{1})x-i\gamma ]\cos [%
\frac{1}{2}(k_{0}-k_{1})x-\alpha ] 
\]%
\begin{equation}
+2\hslash \Omega _{0}(t)I^{-}\exp [-i\frac{1}{2}(k_{0}+k_{1})x+i\gamma ]\cos
[\frac{1}{2}(k_{0}-k_{1})x-\alpha ]  \tag{29}
\end{equation}%
where the electric dipole interaction (23) has been used. In the following
consider a simple case that the on-resonance condition is met, that is, $%
(\omega _{a}-\omega _{0})=0,$ and the amplitude $\Omega _{0}(t)$ is
time-independent, that is, $\Omega _{0}(t)=\Omega _{0}$. Then the
Hamiltonian $H_{I}(t)$ of Eq. (29)\ is reduced to the time-independent form%
\begin{equation}
H_{I}(t)\equiv H_{I}=H_{0}+H_{I}(\alpha ,\gamma )  \tag{30}
\end{equation}%
where $H_{0}$ is still given by Eq. (2) and the electric dipole interaction $%
H_{I}(\alpha ,\gamma )$ is written as 
\[
H_{I}(\alpha ,\gamma )=2\hslash \Omega _{0}I^{+}\exp [i\frac{1}{2}%
(k_{0}+k_{1})x-i\gamma ]\cos [\frac{1}{2}(k_{0}-k_{1})x-\alpha ] 
\]%
\begin{equation}
+2\hslash \Omega _{0}I^{-}\exp [-i\frac{1}{2}(k_{0}+k_{1})x+i\gamma ]\cos [%
\frac{1}{2}(k_{0}-k_{1})x-\alpha ].  \tag{31}
\end{equation}%
Obviously, the Hamiltonian $H_{I}(\alpha ,\gamma )$ is dependent upon the
phases $\alpha $ and $\gamma .$ One may take the time-independent
Hamiltonian $H_{I}$ of Eq. (30) as the basic Hamiltonian to construct the
state-selective trigger pulse. The unitary propagator of the atom system
corresponding to the Hamiltonian (30) is written as 
\begin{equation}
U_{I}(t)=\exp [-iH_{I}t/\hslash ].  \tag{32}
\end{equation}%
On the other hand, in the absence of the two laser light beams the time
evolution process of the atom system is governed by the Hamiltonian $H_{0}$
of Eq. (2) and its unitary propagator is given by%
\begin{equation}
U_{o}(t)=\exp [-iH_{0}t/\hslash ],  \tag{33a}
\end{equation}%
and its inverse propagator by 
\begin{equation}
U_{o}^{+}(t)=\exp [iH_{0}t/\hslash ].  \tag{33b}
\end{equation}%
It can turn out in next section that by using the same Hamiltonian $H_{0}$
of Eq. (2) of the harmonic oscillator one can generate the inverse unitary
propagator $U_{o}^{+}(t)$ with any time $t\neq k\pi /\omega $ up to a global
phase factor. In fact, there are the unitary operator identities: 
\begin{equation}
U_{o}(t)U_{o}(t_{1})=U_{o}(t_{1})U_{o}(t)=\exp [i\beta (t_{1})]E,  \tag{34}
\end{equation}%
where $E$ is the unit operator, $\exp [i\beta (t_{1})]$ is a global phase
factor, and the time $t_{1}=2k\pi /\omega -t$ $(k=1,2,...).$ It follows from
Eqs. (33)\ and (34)\ that one may define the inverse unitary operator $%
U_{o}^{+}(t)$ as%
\begin{equation}
U_{o}^{+}(t)=\exp [-i\beta (t_{1})]U_{o}(t_{1}).  \tag{35}
\end{equation}%
Hereafter the unitary operator $U_{o}(t_{1})$ is also called the inverse
operator of the unitary operator $U_{o}(t)$, although it has a difference of
a global phase factor from the real inverse unitary operator $U_{o}^{+}(t)$.
The unitary propagators $U_{o}(t)$ with any time $t$ of the harmonic
oscillator in the absence of the two laser light beams can be realized
directly in experiment. It follows from Eq. (35)\ that up to a global phase
factor the inverse propagator $U_{o}^{+}(t)$ also can be realized in
experiment as the unitary propagator $U_{o}(t_{1})$ of the harmonic
oscillator can be realized in experiment. On the other hand, the unitary
propagator $U_{I}(t)$ (32) of the harmonic oscillator in the rotating frame
also can be realized directly by the two laser light beams. These three
realizable unitary propagators $U_{o}(t),$ $U_{o}^{+}(t),$ and $U_{I}(t)$ in
the rotating frame are the basic unitary propagators to construct a general
state-selective trigger pulse.

At the first step the unitary propagator of the electric dipole interaction $%
H_{I}(\alpha ,\gamma )$ is created through the three basic unitary
propagators. The unitary propagator of the electric dipole interaction $%
H_{I}(\alpha ,\gamma )$ is defined by 
\begin{equation}
U_{I}(\alpha ,\gamma ,t)=\exp [-iH_{I}(\alpha ,\gamma )t/\hslash ].  \tag{36}
\end{equation}%
It is known from Eq. (30) that the electric dipole interaction $H_{I}(\alpha
,\gamma )$ may be written as $H_{I}(\alpha ,\gamma )=H_{I}-H_{0}.$ Thus, one
may first carry out in experiment a composite pulse sequence:%
\[
U_{o}(t_{1})U_{I}(t)=\exp [i\beta (t_{1})]U_{o}^{+}(t)U_{I}(t) 
\]%
\begin{equation}
=\exp [i\beta (t_{1})]\exp [iH_{0}t/\hslash ]\exp [-iH_{I}t/\hslash ]. 
\tag{37}
\end{equation}%
The physical meaning for the pulse sequence (37) is that a pair of laser
light beams whose parameters satisfy the match condition (22) and time
period is $t$ are first applied to the atom in the harmonic potential well,
then the pair of laser light beams are turned off, and the atom then evolves
in the time period $t_{1}$ in the harmonic potential well without any
external laser light field. If the time interval $\delta t$ is sufficiently
short, then the unitary operator $U_{o}^{+}(\delta t)U_{I}(\delta t)$ is
approximately equal to the unitary propagator $U_{I}(\alpha ,\gamma ,\delta
t)$ according to the famous Baker-Campbell-Hausdorff (BCH) formula for a
product of exponential operators [39, 40, 41], 
\begin{equation}
U_{I}(\alpha ,\gamma ,\delta t)=U_{o}^{+}(\delta t)U_{I}(\delta t)+O((\delta
t)^{2}).  \tag{38}
\end{equation}%
Since the composite pulse sequence $U_{o}(t_{1})U_{I}(t)$ in the rotating
frame can be realized directly in experiment, the composite unitary operator 
$U_{o}^{+}(\delta t)U_{I}(\delta t)$ may be realized up to a global phase
factor, as shown in Eq. (37). Thus, the unitary propagator $U_{I}(\alpha
,\gamma ,\delta t)$ may be realized in experiment up to a global phase
factor, as can be seen from Eq. (38). There is an error term $O((\delta
t)^{2})$ that is proportional to $(\delta t)^{2}$ between the desired
unitary propagator $U_{I}(\alpha ,\gamma ,\delta t)$ and the composite
unitary operator $U_{o}^{+}(\delta t)U_{I}(\delta t)$ on the right-hand side
of Eq. (38). A much better composition for $U_{I}(\alpha ,\gamma ,\delta t)$
with an arbitrary higher-order approximation can also be obtained from the
two propagators $U_{o}^{+}(\delta t)$ and $U(\delta t)$ by using the
Trotter-Suzuki formalism [40, 42]. For example, according to the
Trotter-Suzuki formalism [42] one can obtain a better symmetric composition
for the unitary operator $U_{I}(\alpha ,\gamma ,\delta t)$ by the
multi-pulse sequence: 
\begin{equation}
S_{1}(\delta t)=U_{o}^{+}(\delta t/2)U_{I}(\delta t)U_{o}^{+}(\delta t/2), 
\tag{38a}
\end{equation}%
\[
U_{I}(\alpha ,\gamma ,\delta t)=S_{1}(\delta t)+O((\delta t)^{3}), 
\]%
or a much better symmetric composition by the multi-pulse sequence:%
\begin{equation}
S_{2n-1}(\delta t)=[S_{2n-3}(p_{n}\delta t)]^{2}S_{2n-3}((1-4p_{n})\delta
t)[S_{2n-3}(p_{n}\delta t)]^{2},  \tag{38b}
\end{equation}%
\[
U_{I}(\alpha ,\gamma ,\delta t)=S_{2n-1}(\delta t)+O((\delta t)^{2n+1}), 
\]%
where $n\geq 2$ and $p_{n}=(4-4^{1/(2n-1)})^{-1}.$ Thus, without losing
generality here suppose that the unitary operator $U_{I}(\alpha ,\gamma
,\delta t)$ can be constructed as exactly as pleased. The unitary operator $%
U_{I}(\alpha ,\gamma ,\delta t)$ will be further used to build up the
state-selective trigger pulse.

Now examine the commutation relation between the electric-dipole
interactions $H_{I}(\alpha ,\gamma )$ in two different phases $\gamma =0$
and $\pi /2.$ It can turn out from the electric dipole interaction of Eq.
(31) that the hermite commutation operator $Q$ takes the form%
\begin{equation}
Q\equiv i[H_{I}(\alpha ,0),H_{I}(\alpha ,\pi /2)]=-16\hslash ^{2}\Omega
_{0}^{2}I_{z}\cos ^{2}[\frac{1}{2}(k_{0}-k_{1})x-\alpha ]  \tag{39}
\end{equation}%
where the commutation relations $2I_{z}=[I^{+},I^{-}]$ and $%
[I^{+},I^{+}]=[I^{-},I^{-}]$ $=0$ have been used. If the phase $\alpha =\pi
/4$ and the Lamb-Dicke limit is met, that is, $||(k_{0}-k_{1})x||<<1,$ then
the operator $\cos ^{2}[\frac{1}{2}(k_{0}-k_{1})x-\alpha ]$ can be expanded
as 
\[
\cos ^{2}[\frac{1}{2}(k_{0}-k_{1})x-\alpha ]=\frac{1}{2}+\frac{1}{2}%
(k_{0}-k_{1})x+O(||(k_{0}-k_{1})x||^{3}). 
\]%
Thus, the hermitian operator $Q$ can be written as%
\begin{equation}
Q=-8\hslash ^{2}\Omega _{0}^{2}I_{z}-8\hslash ^{2}(k_{0}-k_{1})\Omega
_{0}^{2}I_{z}x+O(||(k_{0}-k_{1})x||^{3}).  \tag{40}
\end{equation}%
Obviously, the second term on the right-hand side of the operator $Q$ is the
cross term $I_{z}x$ and is responsible for the state-selective excitation of
the coherent state. Therefore, at the second step one should construct the
unitary propagator $\exp (i\lambda Q)$. The unitary operator $\exp (i\lambda
Q)$ can be generated from the unitary operators $\{U_{I}(\alpha ,\gamma
,\delta t)\}$ with the help of the BCH formula [41] and it is realized by
the multi-pulse sequence: 
\[
U_{I}(\alpha ,0,\delta t)U_{I}(\alpha ,\pi /2,\delta t)U_{I}(\alpha
,0,\delta t)^{+}U_{I}(\alpha ,\pi /2,\delta t)^{+} 
\]%
\begin{equation}
=\exp \{-[H_{I}(\alpha ,0),H_{I}(\alpha ,\pi /2)](\delta t/\hslash
)^{2}\}+O((\delta t)^{3}).  \tag{41a}
\end{equation}%
A better composition for the unitary operator $\exp (i\lambda Q)$ can also
be obtained, for example, the following multi-pulse sequence will lead to a
better result, 
\[
\exp [iQ(\delta t/\hslash )^{2}]=U_{I}(\alpha ,0,\delta t/\sqrt{2}%
)U_{I}(\alpha ,\pi /2,\delta t/\sqrt{2}) 
\]%
\[
\times \lbrack U_{I}(\alpha ,0,\delta t/\sqrt{2})^{+}U_{I}(\alpha ,\pi
/2,\delta t/\sqrt{2})^{+}]^{2} 
\]%
\begin{equation}
\times U_{I}(\alpha ,0,\delta t/\sqrt{2})U_{I}(\alpha ,\pi /2,\delta t/\sqrt{%
2})+O((\delta t)^{4}).  \tag{41b}
\end{equation}%
Furthermore, it is possible to obtain a much better composition for the
unitary propagator $\exp [iQ(\delta t/\hslash )^{2}]$ by using the
Trotter-Suzuki formalism [42, 43]. Here it will not be further discussed in
detail. By neglecting the error term $O(||(k_{0}-k_{1})x||^{3})$ on the
right-hand side of Eq. (40) the unitary operator $\exp [iQ(\delta t/\hslash
)^{2}]$ is written as 
\begin{equation}
\exp [iQ(\delta t/\hslash )^{2}]=\exp \{-i8\Omega _{0}^{2}[I_{z}+\Delta
kI_{z}x](\delta t)^{2}\}.  \tag{42}
\end{equation}%
It can turn out below that the unitary propagator $\exp [iQ(\delta t/\hslash
)^{2}]$ can excite directly and state-selectively the ground motional state
of the harmonic oscillator to a Gaussian wave-packet motional state with a
high motional energy.

Now suppose that at the initial time the atom is in the product state $\exp
(i\varphi _{0})$ $\times |\psi _{0}(x)\rangle |g_{1}\rangle $ in the
rotating frame, where $\exp (i\varphi _{0})$ is a global phase factor.
Applying the unitary propagator $\exp [iQ(\delta t/\hslash )^{2}]$ to the
initial product state one obtains 
\begin{equation}
|\Psi (x,r,t)\rangle =\exp [iQ(\delta t/\hslash )^{2}]\exp (i\varphi
_{0})|\psi _{0}(x)\rangle |g_{1}\rangle =\exp (i\varphi _{0})|\psi
_{0}(x)\rangle |g_{1}\rangle  \tag{43}
\end{equation}%
where the eigen-equation $I_{z}|g_{1}\rangle =0|g_{1}\rangle $ has been
used. Thus, the initial product state keeps unchanged under the unitary
propagator $\exp [iQ(\delta t/\hslash )^{2}].$ On the other hand, if the
initial product state of the atom is $\exp (i\varphi _{0})|\psi
_{0}(x)\rangle |g_{0}\rangle $ or $\exp (i\varphi _{0})|\psi _{0}(x)\rangle
|e\rangle $ in the rotating frame, then applying the unitary propagator to
the initial product state one obtains the atomic product state 
\begin{equation}
|\Psi (x,r,t)\rangle =\exp (i\varphi _{0})\exp \{i4\Omega _{0}^{2}(\delta
t)^{2}\}\exp \{i4\Omega _{0}^{2}\Delta k(\delta t)^{2}x\}|\psi
_{0}(x)\rangle |g_{0}\rangle .  \tag{43a}
\end{equation}%
or%
\begin{equation}
|\Psi (x,r,t)\rangle =\exp (i\varphi _{0})\exp \{-i4\Omega _{0}^{2}(\delta
t)^{2}\}\exp \{-i4\Omega _{0}^{2}\Delta k(\delta t)^{2}x\}|\psi
_{0}(x)\rangle |e\rangle .  \tag{43b}
\end{equation}%
where the eigen-equations $I_{z}|g_{0}\rangle =(-1/2)|g_{0}\rangle $ and $%
I_{z}|e\rangle =(1/2)|e\rangle $ have been used. If now the rotating frame
is changed back to the laboratory frame, then only a global phase factor is
generated for each of the three atomic product states $|\Psi (x,r,t)\rangle $
of Eqs. (43), (43a), and (43b). Thus, the atomic product states $|\Psi
(x,r,t)\rangle $ of Eqs. (43), (43a), and (43b) are really the final atomic
product states after the atom is applied by the double-frequency pulse
sequence that generates the unitary propagator $\exp [iQ(\delta t/\hslash
)^{2}]$.\ Now the ground motional state $|\psi _{0}(x)\rangle $ of the
harmonic oscillator takes the Gaussian form 
\begin{equation}
|\psi _{0}(x)\rangle =[\frac{1}{2\pi (\Delta x)_{0}^{2}}]^{1/4}\exp [-\frac{1%
}{4}\frac{x^{2}}{(\Delta x)_{0}^{2}}]  \tag{44}
\end{equation}%
where $(\Delta x)_{0}^{2}=(\frac{\hslash }{2m\omega })$ and $\omega $ is the
oscillatory frequency of the harmonic oscillator. Obviously, this motional
state keeps unchanged during the unitary propagator $\exp [iQ(\delta
t/\hslash )^{2}]$ acting on the atom when the atom is in the internal state $%
|g_{1}\rangle ,$ as shown in Eq. (43). However, the atomic product states
(43a)\ and (43b)\ show that after the unitary operator $\exp [iQ(\delta
t/\hslash )^{2}]$ acts on the initial product state $\exp (i\varphi
_{0})|\psi _{0}(x)\rangle |g_{0}\rangle $ or $\exp (i\varphi _{0})|\psi
_{0}(x)\rangle |e\rangle $ the atom is in the motional state 
\begin{equation}
|\Psi (x,t)\rangle =\exp [i\varphi (t)][\frac{1}{2\pi (\Delta x)_{0}^{2}}%
]^{1/4}\exp [-\frac{1}{4}\frac{x^{2}}{(\Delta x)_{0}^{2}}]\exp
[ip_{0}x/\hslash ]  \tag{44a}
\end{equation}%
or%
\begin{equation}
|\Psi (x,t)\rangle =\exp [i\phi (t)][\frac{1}{2\pi (\Delta x)_{0}^{2}}%
]^{1/4}\exp [-\frac{1}{4}\frac{x^{2}}{(\Delta x)_{0}^{2}}]\exp
[-ip_{0}x/\hslash ]  \tag{44b}
\end{equation}%
where the global phase factor $\exp [i\varphi (t)]=\exp (i\varphi _{0})\exp
\{i4\Omega _{0}^{2}(\delta t)^{2}\},$ $\exp [i\phi (t)]=\exp (i\varphi
_{0})\exp \{-i4\Omega _{0}^{2}(\delta t)^{2}\},$ and the mean momentum $%
p_{0} $ is given by $p_{0}/\hslash =4\Omega _{0}^{2}\Delta k(\delta t)^{2}.$
Both the motional states (44a) and (44b)\ are the Gaussian wave-packet
states with the center-of-mass positions $x_{0}=0$ and the momentums $p_{0}$
and $-p_{0},$ respectively. Thus, the two motional states show that after
the double-frequency pulse sequence is turned off the atom is still in the
original harmonic potential well and has approximately the mean motional
energy, 
\begin{equation}
E_{0}=p_{0}^{2}/(2m)+\frac{1}{2}m\omega ^{2}x_{0}^{2}=[4\hslash \Delta
k\Omega _{0}^{2}(\delta t)^{2}]^{2}/(2m).  \tag{45}
\end{equation}%
This motional energy $E_{0}$ can be much larger than the zero-point energy
of the atom in the harmonic potential well. On the other hand, the motional
state (44a) shows that at the end of the excitation process of the unitary
propagator $\exp [iQ(\delta t/\hslash )^{2}]$ the atom in the internal state 
$|g_{0}\rangle $ moves along the direction $+x$ with the velocity $p_{0}/m$
in the harmonic potential well, while the motional state (44b) shows that
the atom in the internal state $|e\rangle $ moves along the direction $-x$
with the same velocity $p_{0}/m$. The above investigation for the effect of
the unitary propagator $\exp [iQ(\delta t/\hslash )^{2}]$ on the initial
product state of the atom is summarized as follows. When the atom is in the
internal state $|g_{1}\rangle $ at the initial time$,$ the unitary
propagator $\exp [iQ(\delta t/\hslash )^{2}]$ does not have any significant
effect on the initial product state. In particular, even if the initial
motional state of the atom is an arbitrary wave function, i.e., a
superposition motional state, the unitary propagator $\exp [iQ(\delta
t/\hslash )^{2}]$ does not yet have any siginificant effect on the motional
state when the atom is in the internal state $|g_{1}\rangle $ at the initial
time. However, the initial ground motional state of the atom will be
transferred to a Gaussian wave-packet state with a higher motional energy by
the unitary propagator if the atom is in the internal state $|g_{0}\rangle $
or $|e\rangle $ at the initial time. Therefore, the double-frequency pulse
sequences (41) to generate the unitary propagator $\exp [iQ(\delta t/\hslash
)^{2}]$ may generally act as the state-selective trigger pulse in the
quantum control process [1].

In order to implement the double-frequency pulse sequences (41) one needs to
realize not only the unitary propagators $\{U_{I}(\alpha ,\gamma ,t)\}$ but
also their inverse propagators $\{U_{I}^{+}(\alpha ,\gamma ,t)\}.$ The
inverse unitary operators $\{U_{I}^{+}(\alpha ,\gamma ,\delta t)\}$ can be
generated as follows. Since the electric dipole Hamiltonian $H_{I}(\alpha
,\gamma )$ of Eq. (31)\ is time-independent, these inverse unitary operators
may be generated by inverting the electric dipole Hamiltonian: $H_{I}(\alpha
,\gamma )\rightarrow -H_{I}(\alpha ,\gamma ).$ Since the electric dipole
Hamiltonian is dependent upon the phases $\alpha $ and $\gamma ,$ it is
possible to choose suitably the phase $\gamma $ to obtain the negative-sign
Hamiltonian $-H_{I}(\alpha ,\gamma ).$ In fact, it follows from Eq. (31)
that $H_{I}(\alpha ,0)=-H_{I}(\alpha ,\pi )$ and $H_{I}(\alpha ,\pi
/2)=-H_{I}(\alpha ,3\pi /2).$ Therefore, the inverse unitary operators $%
U_{I}^{+}(\alpha ,0,\delta t)$ and $U_{I}^{+}(\alpha ,\pi /2,\delta t)$ are
respectively given by 
\begin{equation}
U_{I}^{+}(\alpha ,0,\delta t)=U_{I}(\alpha ,\pi ,\delta t),\text{ }%
U_{I}^{+}(\alpha ,\pi /2,\delta t)=U_{I}(\alpha ,3\pi /2,\delta t).  \tag{46}
\end{equation}%
Note that the unitary propagator $U_{I}(\alpha ,\gamma ,\delta t)$ for any
phase values $\alpha $ and $\gamma $ can be implemented in experiment. Then
the unitary propagator $\exp [iQ(\delta t/\hslash )^{2}]$ can be implemented
in experiment, since it follows from Eqs. (41b) and (46)\ that the unitary
propagator $\exp [iQ(\delta t/\hslash )^{2}]$ can be expressed as 
\[
\exp [iQ(\delta t/\hslash )^{2}]=U_{I}(\alpha ,0,\delta t/\sqrt{2}%
)U_{I}(\alpha ,\pi /2,\delta t/\sqrt{2})U_{I}(\alpha ,\pi ,\delta t/\sqrt{2}%
) 
\]%
\[
\times U_{I}(\alpha ,3\pi /2,\delta t/\sqrt{2})U_{I}(\alpha ,\pi ,\delta t/%
\sqrt{2})U_{I}(\alpha ,3\pi /2,\delta t/\sqrt{2}) 
\]%
\begin{equation}
\times U_{I}(\alpha ,0,\delta t/\sqrt{2})U_{I}(\alpha ,\pi /2,\delta t/\sqrt{%
2})+O((\delta t)^{4}).  \tag{47}
\end{equation}%
The double-frequency pulse sequence (47)\ of the propagator $\exp [iQ(\delta
t/\hslash )^{2}]$ consists of a number of the three realizable unitary
propagators $U_{o}(t),$ $U_{o}^{+}(t),$ and $U_{I}(t).$ Obviously, it can be
efficiently implemented in experiment, as can be seen from Eqs. (37), (38),
(38a), and (47). It should be pointed out that one also can obtain a much
better multi-pulse sequence than the sequence (47) for the unitary
propagator $\exp [iQ(\delta t/\hslash )^{2}]$ with the help of the
Trotter-Suzuki formalism [42, 43]. This fact tells ones that the unitary
propagator $\exp [iQ(\delta t/\hslash )^{2}]$ can be implemented efficiently
and as exactly as pleased.

On the other hand, the internal-state rotating operator $R_{z}(\theta )=\exp
(-i\theta I_{z})$ of the atom in the harmonic potential well in the rotating
frame can also be prepared in a similar way to generating the unitary
propagator $\exp [iQ(\delta t/\hslash )^{2}].$ When the phase $\alpha =0$
and the Lamb-Dicke limit $||(k_{0}-k_{1})x||<<1$ is met, the hermitian
operator $Q$ of Eq. (39)\ may be written as 
\[
Q\equiv -16\hslash ^{2}\Omega _{0}^{2}I_{z}+O(||(k_{0}-k_{1})x||^{2}). 
\]%
If the error term $O(||(k_{0}-k_{1})x||^{2})$ is neglected, then the unitary
propagator $\exp [iQ(\delta t/\hslash )^{2}]$ is really the internal-state
rotating operator:%
\begin{equation}
R_{z}(\theta )=\exp [iQ(\delta t/\hslash )^{2}]=\exp \{-i16\Omega
_{0}^{2}(\delta t)^{2}I_{z}\}  \tag{48}
\end{equation}%
where the rotating angle $\theta =16\Omega _{0}^{2}(\delta t)^{2}.$ This
rotating operator is independent of any atomic motional state but applied
only to the two atomic internal states $|g_{0}\rangle $ and $|e\rangle $
selectively. Here it must be pointed out that the inverse propagator of the
unitary propagator $\exp [iQ(\delta t/\hslash )^{2}]$ and the inverse
operator of the internal-state rotating unitary operator $R_{z}(\theta )$
can also be implemented in experiment, as can be seen from Eqs. (41) and
(47), because both the unitary operator $U_{I}(\alpha ,\gamma ,t)$ and its
inverse operator $U_{I}^{+}(\alpha ,\gamma ,t)$ can be implemented in
experiment.

In this section the on-resonance condition $\omega _{a}=\omega _{0}$ has
been used in the Hamiltonian (29) to simplify the construction of the
propagator $\exp [iQ(\delta t/\hslash )^{2}].$ For a general case that the
on-resonance condition does not hold, that is, $\omega _{a}\neq \omega _{0},$
one may use $\pi $ pulses $\exp (-i\pi I_{x})$ and/or $\exp (-i\pi I_{y})$
to refocus the term $\hslash (\omega _{a}-\omega _{0})I_{z}$ in the
Hamiltonian (29), where the $\pi $ pulses $\exp (-i\pi I_{x})$ and $\exp
(-i\pi I_{y})$ may be generated by an ultrashort laser light pulse. Then in
the general case the coherent average method can be used as well to build up
the unitary propagator $\exp [iQ(\delta t/\hslash )^{2}].$ The programming
operator composition method has also been used to build up the quantum gates
and/or the internal-state-selective quantum gates in the trapped ion systems
[44, 45, 46].\newline
\newline
\newline
{\large 3. Manipulating the complex linewidth of a Gaussian wave-packet state%
}

The Gaussian wave-packet motional state or the standard coherent state
generated by the state-selective trigger pulse must have a higher motional
energy than the ground state of the harmonic oscillator, but the Gaussian
wave-packet state generated by the state-selective trigger pulse could not
have an expected complex linewidth. However, in the quantum control process
to realize the reversible and unitary state-insensitive halting protocol and
the efficient quantum search process it could be required that the Gaussian
wave-packet state of the halting-qubit atom have an adjustable complex
linewidth after the state-selective trigger pulse really acts on the atom.
Thus, one needs to construct a pulse sequence to control the complex
linewidth of a Gaussian wave-packet state. This pulse sequence combining
with the state-selective trigger pulse will form a composite state-selective
trigger pulse. This composite state-selective trigger pulse may manipulate
not only the center-of-mass position and momentum but also the complex
linewidth of the Gaussian wave-packet state when the halting-qubit atom is
really acted on by the composite state-selective trigger pulse. Below it is
discussed how to generate a pulse sequence to manipulate the complex
linewidth of a Gaussian wave-packet state. The complex linewidth for a
Gaussian wave-packet state could be controlled by an external harmonic
potential field. It is known that the initial motional state of the
halting-qubit atom is prepared to be the ground state of the harmonic
oscillator, which is the Gaussian wave-packet state $|\psi _{0}(x)\rangle $
of Eq. (44), and the harmonic potential field that is applied to the atom
has the oscillatory frequency $\omega .$ Now this harmonic potential field
is switched to another harmonic potential field with the oscillatory
frequency $\omega _{c}$ at the initial time. The atom then undergoes a time
evolution process in the new harmonic potential field. This evolution
process may be described by the unitary propagator of the harmonic
oscillator with the oscillatory frequency $\omega _{c}$, which in the
coordinate representation may be expressed as [13, 14, 21]%
\begin{equation}
G(x_{b},t_{b};x_{a},t_{a})=\sqrt{\frac{m\omega _{c}}{i2\pi \hslash }}\exp [-i%
\frac{m\omega _{c}}{\hslash }x_{a}x_{b}]  \tag{49}
\end{equation}%
where the period of the process $T_{c}=t_{b}-t_{a}$ is chosen such that it
satisfies $\cos [\omega _{c}T_{c}]=0$ and $\sin [\omega _{c}T_{c}]=1,$ that
is, $T_{c}=[2k\pi +\pi /2]/\omega _{c}$ ($k=0,1,...$). The atomic wave
function at the end of the process is calculated by 
\[
\Psi (x_{b},t_{b})=\int dx_{a}G(x_{b},t_{b};x_{a},t_{a})\psi _{0}(x_{a}) 
\]%
\[
=[\frac{1}{2\pi (\Delta x)_{0}^{2}}]^{1/4}\sqrt{\frac{m\omega _{c}}{i2\pi
\hslash }}\int dx_{a}\exp (-ax_{a}^{2}+bx_{a}), 
\]%
where the ground state $\psi _{0}(x_{a})$ of Eq. (44)\ is used and the
parameter $a=[4(\Delta x)_{0}^{2}]^{-1}$ and $b=-i\hslash ^{-1}m\omega
_{c}x_{b}.$ By using the Gaussian integral formula: 
\begin{equation}
\int dx_{a}\exp (-ax_{a}^{2}+bx_{a})=\sqrt{\frac{\pi }{a}}\exp (\frac{b^{2}}{%
4a})  \tag{50}
\end{equation}%
one obtains the wave function:%
\begin{equation}
\Psi (x_{b},t_{b})=\exp (i\phi _{0})[\frac{1}{2\pi (\Delta x)^{2}}%
]^{1/4}\exp \{-\frac{1}{4}\frac{x_{b}^{2}}{(\Delta x)^{2}}\},  \tag{51}
\end{equation}%
where the phase $\phi _{0}=-\pi /4$ and the wave-packet spreading is just $%
\varepsilon =\sqrt{2}(\Delta x),$ and $(\Delta x)^{2}$ is given by 
\[
(\Delta x)^{2}=[\frac{\hslash }{2m\omega _{c}(\Delta x)_{0}}]^{2}=(\frac{%
\omega }{\omega _{c}})(\frac{\hslash }{2m\omega _{c}}) 
\]%
where $(\Delta x)_{0}^{2}=(\frac{\hslash }{2m\omega })$ is used. The
imaginary part of the complex linewidth of the Gaussian wave-packet state
(51)\ is zero and the real part is $(\Delta x)^{2}.$ Therefore, the
wave-packet spreading $\varepsilon $ or the real part of the complex
linewidth is controlled by both the oscillatory frequencies $\omega _{c}$
and $\omega $ of the harmonic potential fields. The initial wave-packet
motional state $\psi _{0}(x)$ (44)\ usually may be prepared to have a small
and fixed wave-packet spreading $\varepsilon _{0}=\sqrt{2}(\Delta x)_{0}$
which corresponds to a large oscillatory frequency $\omega .$ For example,
if one takes $\omega \thicksim 10^{8},$ then $(\Delta x)_{0}^{2}\thicksim
10^{-17}$ and $\omega \hslash /(2m)\thicksim 0.1$ for the atomic mass $%
m\thicksim 10^{-25}$Kg. Then by setting suitably the oscillatory frequency $%
\omega _{c}$ one may obtain the desired wave-packet spreading or the real
part of the complex linewidth for the state $\Psi (x_{b},t_{b})$ (51). The
important thing is that the center-of-mass position and momentum for the
Gaussian wave-packet state of the atom keeps unchanged in the evolution
process, as can be seen from the state $\psi _{0}(x)$ (44)\ and the state $%
\Psi (x_{b},t_{b})$ (51). This means that after the evolution process the
atom is still in the original position $x=0$ in the coordinate axis and has
zero momentum. One therefore concludes that the real part of the complex
linewidth of a Gaussian wave-packet state of an atom may be manipulated by
varying the oscillatory frequency of the harmonic potential field applying
to the atom.

The imaginary part of the complex linewidth of a Gaussian wave-packet state
may also be manipulated at will. One of the simplest and most intuitive
methods to manipulate the imaginary part of the complex linewidth is that
the atom undergoes simply a free-particle motion or an inverse free-particle
motion. It is well known that the wave-packet spreading of the Gaussian
wave-packet state of a free atom (the initial imaginary part of the complex
linewidth is zero) becomes larger and larger when the atom undergoes a
free-particle motion [2]. Obviously, the imaginary part of the complex
linewidth will become less and less if the atom undergoes the inverse
free-particle motion. Suppose that at the initial time the atom in the
harmonic potential well is in the Gaussian wave-packet state of Eq. (51).
Now one turns off the harmonic potential field applying to the atom. Then
the atom becomes a free atom. However, the atom does not leave its original
position even after the harmonic potential field is switched off, since the
atomic motional momentum is zero before the harmonic potential field is
turned off. Therefore, the only change for the Gaussian wave-packet state
(51) of the free atom is its complex linewidth after the harmonic potential
field is turned off. The time evolution process of the atom after the
harmonic potential field is turned off may be calculated by using the
free-particle unitary propagator. It is known that the unitary propagator of
a free particle is given by [13, 14, 21]%
\begin{equation}
G(x_{b},t_{b};x_{a},t_{a})=\sqrt{\frac{m}{i2\pi hT}}\exp [i\frac{m}{2\hslash
T}(x_{b}-x_{a})^{2}]  \tag{52a}
\end{equation}%
where $T=t_{b}-t_{a}$ is the time period of the free-particle motion. After
the harmonic potential field is turned off the free atom undergoes a
free-particle motion with the time period $T$ and its motional state $\Psi
(x_{a},t_{a})$ of Eq. (51) is changed to another Gaussian wave-packet state: 
\begin{equation}
\Psi _{f}(x_{b},t_{b})=\exp (i\phi _{0})[\frac{1}{2\pi (\Delta x)^{2}}%
]^{1/4}\exp \{-\frac{1}{4}\frac{x_{b}^{2}}{(\Delta x)^{2}+i(\frac{\hslash T}{%
2m})}\}.  \tag{53a}
\end{equation}%
Here the complex linewidth of the Gaussian wave-packet state $\Psi
_{f}(x_{b},t_{b})$ is given by $W(T)=(\Delta x)^{2}+i\hslash T/(2m).$ Thus,
the imaginary part of the complex linewidth is proportional to the time
period $T$ of the free-particle motion. However, the imaginary part $\hslash
T/(2m)$ is always positive. In order to achieve a negative imaginary part
one may let the atom perform an inverse free-particle motion. The unitary
propagator for the inverse free-particle motion may be given by%
\begin{equation}
G^{+}(x_{b},t_{b};x_{a},t_{a})=\sqrt{-\frac{m}{i2\pi \hslash T}}\exp [-i%
\frac{m}{2\hslash T}(x_{b}-x_{a})^{2}]  \tag{52b}
\end{equation}%
The unitary propagator of the inverse free-particle motion may be generated
with the help of the external quadratic potential field (see below). Now the
state $\Psi (x_{a},t_{a})$ of Eq. (51) is changed to the Gaussian
wave-packet state $\Psi _{i}(x_{b},t_{b})$ after the inverse free-particle
motion, 
\begin{equation}
\Psi _{i}(x_{b},t_{b})=\exp [i\varphi _{0}][\frac{1}{2\pi (\Delta x)^{2}}%
]^{1/4}\exp \{-\frac{1}{4}\frac{x_{b}^{2}}{(\Delta x)^{2}-i(\frac{\hslash T}{%
2m})}\}.  \tag{53b}
\end{equation}%
Here the complex linewidth of the state $\Psi _{i}(x_{b},t_{b})$ is given by 
$W=(\Delta x)^{2}-i\hslash T/(2m).$ Its imaginary part is negative. Thus,
the imaginary part of the complex linewidth can be controlled by the
free-particle motion and the inverse free-particle motion. Obviously, due to
the fact that the atomic motional momentum is zero the center-of-mass
position ($x_{0}=0$) of the Gaussian wave-packet state (51) keeps unchanged
in the free-particle motion and the inverse free-particle motion. The
manipulating method by using the free-particle motion and the inverse
free-particle motion is very simple. However, there could be a disadvantage
for the manipulation: if a large imaginary part of the complex linewidth
needs to be achieved, then one needs to spend a long time for it.

In the preceding discussion one needs to use the inverse unitary propagator
of a free particle to manipulate the imaginary part of the complex
linewidth. Here gives the explicit expression for the inverse unitary
propagator without a detail proof. It is known that the unitary propagator
of a free particle is given by $U_{f}(t)=\exp [-\frac{p^{2}t}{2m\hslash }],$ 
$p$ is the momentum operator of the free particle. Denote that $%
U_{ok}(t_{k})=\exp [-iH_{0k}t_{k}/\hslash ]$ ($k=1,2$) is a unitary
propagator of a harmonic oscillator. The Hamiltonian $H_{0k}$ of the
harmonic oscillator is $H_{0k}=p^{2}/(2m)+m\omega _{k}^{2}x^{2}/2.$ Then it
can prove that the inverse unitary propagator of a free particle can be
written as, up to a global phase factor, 
\[
U_{f}(T)^{+}\equiv \exp [i\frac{p^{2}T}{2m\hslash }%
]=U_{o1}(T_{1})U_{f}(T)U_{o2}(T_{2}) 
\]%
where the time intervals $T_{1}$ and $T_{2}$ (or the oscillatory frequencies 
$\omega _{1}$ and $\omega _{2}$ of the two harmonic potential fields) are
determined through 
\[
\sin (\omega _{1}T_{1})=\mp \frac{2T\omega _{1}\omega _{2}^{2}}{\sqrt{%
[T^{2}\omega _{1}^{2}\omega _{2}^{2}-(\omega _{2}^{2}-\omega
_{1}^{2})]^{2}+[2T\omega _{1}\omega _{2}^{2}]^{2}}}, 
\]%
\[
\cos (\omega _{1}T_{1})=\mp \frac{\lbrack T^{2}\omega _{1}^{2}\omega
_{2}^{2}-(\omega _{2}^{2}-\omega _{1}^{2})]}{\sqrt{[T^{2}\omega
_{1}^{2}\omega _{2}^{2}-(\omega _{2}^{2}-\omega _{1}^{2})]^{2}+[2T\omega
_{1}\omega _{2}^{2}]^{2}}} 
\]%
and 
\[
\sin (\omega _{2}T_{2})=\pm \frac{2T\omega _{1}^{2}\omega _{2}}{\sqrt{%
[T^{2}\omega _{1}^{2}\omega _{2}^{2}+(\omega _{2}^{2}-\omega
_{1}^{2})]^{2}+[2T\omega _{1}^{2}\omega _{2}]^{2}}}, 
\]%
\[
\cos (\omega _{2}T_{2})=\pm \frac{\lbrack T^{2}\omega _{1}^{2}\omega
_{2}^{2}+(\omega _{2}^{2}-\omega _{1}^{2})]}{\sqrt{[T^{2}\omega
_{1}^{2}\omega _{2}^{2}+(\omega _{2}^{2}-\omega _{1}^{2})]^{2}+[2T\omega
_{1}^{2}\omega _{2}]^{2}}}. 
\]%
Given the time period $T$ of the free-particle motion and the oscillatory
frequencies $\omega _{1}$ and $\omega _{2}$ of the two harmonic potential
fields one can calculate the time intervals $T_{1}$ and $T_{2}$ from the
four equations above. One sees that the inverse unitary propagator of a free
particle can be realized only when the specific external harmonic potential
fields are applied to the particle. Therefore, the time evolution process of
a free particle is a self-irreversible evolution process, although this
process is unitary.

To improve the simple method based on the free-particle motion or the
inverse free-particle motion more complex multi-pulse sequences may be
employed to adjust the complex linewidth. The pulse sequences consist of
several pulses of the harmonic potential fields with different oscillatory
frequencies. One of the pulse sequences is given below. It is known that
after the free-particle motion with a short time period $T$ the atom is in
the Gaussian wave-packet state of Eq. (53a). Now the atom is applied by a
pulse sequence consisting of two harmonic potential field pulses with
different oscillatory frequencies (see below) such that the time evolution
process of the atom is described by the unitary propagator: 
\begin{equation}
G(x_{b},t_{b},x_{a},t_{a})=\sqrt{\frac{m(-S_{ab})}{i2\pi \hslash }}\exp \{i(%
\frac{m}{2\hslash })[S_{bb}x_{b}^{2}+S_{ab}2x_{a}x_{b}]\}  \tag{54}
\end{equation}%
where these parameters in the propagator are given later. Now one has the
initial state of Eq. (53a) with the\ time parameter $T=T_{0}$ and the
propagator of Eq. (54). Then one can determine the time evolution process of
the atom under the harmonic-potential-field pulse sequence. The final state
of the process is given by%
\[
\Psi (x_{b},t_{b})=\exp (i\phi _{0})[\frac{(\Delta x)^{2}}{2\pi }]^{1/4}%
\sqrt{\frac{2m(-S_{ab})}{i\hslash }} 
\]%
\begin{equation}
\times \exp \{-\frac{1}{4}(\frac{2m}{\hslash })[(\frac{2m(\Delta x)^{2}}{%
\hslash })S_{ab}^{2}+i(T_{0}S_{ab}^{2}-S_{bb})]x_{b}^{2}\}.  \tag{55}
\end{equation}%
The Gaussian wave-packet state $\Psi (x_{b},t_{b})$ (55)\ has the complex
linewidth: \newline
\begin{equation}
W=(\frac{\hslash }{2m})\frac{\delta S_{ab}^{2}-i(T_{0}S_{ab}^{2}-S_{bb})}{%
\delta ^{2}S_{ab}^{4}+(T_{0}S_{ab}^{2}-S_{bb})^{2}}  \tag{56}
\end{equation}%
where $\delta =2m(\Delta x)^{2}/\hslash .$ Denote the complex linewidth $%
W=(\Delta y)^{2}+i\hslash T_{w}/(2m).$ Then one has%
\begin{equation}
T_{w}=-\frac{(T_{0}S_{ab}^{2}-S_{bb})}{\delta
^{2}S_{ab}^{4}+(T_{0}S_{ab}^{2}-S_{bb})^{2}}  \tag{57}
\end{equation}%
and 
\begin{equation}
(\Delta y)^{2}=\frac{(\Delta x)^{2}S_{ab}^{2}}{\delta
^{2}S_{ab}^{4}+(T_{0}S_{ab}^{2}-S_{bb})^{2}}.  \tag{58}
\end{equation}%
Obviously, $T_{w}\geq 0$ if $(T_{0}S_{ab}^{2}-S_{bb})\leq 0$ and $T_{w}<0$
if $(T_{0}S_{ab}^{2}-S_{bb})>0.$ Only when $\delta
^{2}S_{ab}^{4}+(T_{0}S_{ab}^{2}-S_{bb})^{2}<<1$ can it be possible for the
absolute parameter $|T_{w}|$ to be much larger than one. The equations (57)
and (58) show that both the real ($(\Delta y)^{2}$) and imaginary ($T_{w}$)
parts of the complex linewidth can be controlled simultaneously by the
harmonic-potential-field pulse sequence.

Suppose that the harmonic-potential-field pulse sequence is applied to the
atom in the manner that the first harmonic potential field with the
oscillatory frequency $\omega $ is applied to the atom at the initial time,
it lasts a time interval $T,$ then it is turned off and at the same time the
second harmonic potential field with the oscillatory frequency $\omega _{o}$
is turned on, and then it lasts a time interval $T_{o}.$ Thus, the total
time period of the harmonic-potential-field pulse sequence is $T+T_{o}.$ It
is known that the unitary propagator of a harmonic oscillator is generally
written as [13, 14, 21], 
\begin{equation}
G(x_{b},t_{b};x_{a},t_{a})=\sqrt{\frac{m\omega }{i2\pi \hslash \sin (\omega
T)}}\exp \{i\frac{m\omega }{2\hslash \sin (\omega T)}[(x_{b}^{2}+x_{a}^{2})%
\cos (\omega T)-2x_{b}x_{a}]\}  \tag{59}
\end{equation}%
The propagator (49) is a special case of the propagator (59)\ with the time
period $T=[2k\pi +\pi /2]/\omega .$ One can see that when the time period $T$
satisfies $\omega T=k\pi ,$ the propagator (59)\ appears singular as $\sin
(\omega T)=0.$ It seems that the propagator of the harmonic oscillator\
should take a different form from the original one (59) at these time period
points that make the propagator singular. But it can turn out that even at
these time period points the correct propagator of the harmonic oscillator
can be obtained directly from the original one (59) [47, 48]. Therefore, the
unitary propagator (59) indeed can describe completely the time evolution
process of a harmonic oscillator with any time period. The composite unitary
propagator of the harmonic-potential-field pulse sequence then is calculated
by 
\begin{equation}
G(x_{b},t_{b};x_{a},t_{a})=\int
dx_{c}G_{2}(x_{b},t_{b};x_{c},t_{c})G_{1}(x_{c},t_{c};x_{a},t_{a})  \tag{60}
\end{equation}%
where $\{G_{k}(x^{\prime },t^{\prime };x,t),$ $k=1,$ $2\}$ are the unitary
propagators of the atom under the first ($k=1$) and the second ($k=2$)\
harmonic potential field, respectively. Note that $T=t_{c}-t_{a}$, $%
T_{o}=t_{b}-t_{c},$ and $t_{b}-t_{a}=T+T_{o}.$ By substituting the unitary
propagator (59) of the harmonic oscillator in the equation (60) and then by
a complex calculation one obtains 
\begin{equation}
G(x_{b},t_{b},x_{a},t_{a})=\sqrt{(\frac{m\omega \omega _{o}}{i2\pi \hslash
\eta })}\exp \{i(\frac{m}{2\hslash }%
)[S_{bb}x_{b}^{2}+S_{ab}2x_{a}x_{b}+S_{aa}x_{a}^{2}]\}  \tag{61}
\end{equation}%
where the parameters are given by 
\[
\eta =[\omega _{o}\cos (\omega _{o}T_{o})\sin (\omega T)+\omega \sin (\omega
_{o}T_{o})\cos (\omega T)], 
\]%
\[
S_{bb}=\frac{\omega _{o}}{\eta }[-\omega _{o}\sin (\omega T)\sin (\omega
_{o}T_{o})+\omega \cos (\omega _{o}T_{o})\cos (\omega T)], 
\]%
\[
S_{ab}=-\frac{\omega _{o}\omega }{\eta }, 
\]%
\[
S_{aa}=\frac{\omega }{\eta }[-\omega \sin (\omega T)\sin (\omega
_{o}T_{o})+\omega _{o}\cos (\omega T)\cos (\omega _{o}T_{o})]. 
\]%
If one sets the parameter $S_{aa}=0$ in Eq. (61), then the propagator (61)
is reduced to the propagator (54). Since the parameter $\eta $ satisfies $%
0<|\eta |\leq \omega +\omega _{o},$ the parameter $S_{aa}=0$ means that 
\begin{equation}
\tan (\omega T)\tan (\omega _{o}T_{o})=\frac{\omega _{o}}{\omega }.  \tag{62}
\end{equation}%
By using the equation (62) one can reduce respectively the parameters $%
S_{bb} $ and $S_{ab}^{2}$ to the forms: 
\[
S_{bb}=\frac{(\omega ^{2}-\omega _{o}^{2})}{\omega }\frac{\tan (\omega T)}{%
[1+\tan ^{2}(\omega T)]}, 
\]%
\[
S_{ab}^{2}=\frac{\omega _{o}^{2}+\omega ^{2}\tan ^{2}(\omega T)}{1+\tan
^{2}(\omega T)}. 
\]%
From these parameters one sees that both the wave-packet spreading $\sqrt{2}%
(\Delta y)^{2}$ (58) and the time interval $T_{w}$ (57)\ are dependent upon
the parameters $\omega ,$ $T,$ $\omega _{o},$ and $T_{o}$ of the two
harmonic potential fields. There are four undetermined parameters $\omega ,$ 
$T,$ $\omega _{o},$ and $T_{o}$ of the two harmonic potential fields, while
there are only three independent equations (57), (58), and (62) to determine
these parameters. Thus, given the complex linewidth $W=(\Delta
y)^{2}+i\hslash T_{w}/(2m)$ for the Gaussian wave-packet state $\Psi
(x_{b},t_{b})$ (55), one can determine these parameters for the two harmonic
potential fields.

The equations (57) and (58) lead to the relation:%
\begin{equation}
\frac{(\Delta x)^{2}T_{w}}{(\Delta y)^{2}}=-\frac{(T_{0}S_{ab}^{2}-S_{bb})}{%
S_{ab}^{2}}.  \tag{63}
\end{equation}%
Substituting the parameters $S_{bb}$ and $S_{ab}^{2}$ in the equation (63)
one obtains 
\begin{equation}
\frac{\omega T_{w}(\Delta x)^{2}}{(\Delta y)^{2}}=-\frac{(\omega
T_{0})[n_{o}^{2}+\tan ^{2}(\omega T)]-(1-n_{o}^{2})\tan (\omega T)}{%
[n_{o}^{2}+\tan ^{2}(\omega T)]}  \tag{64}
\end{equation}%
where $n_{o}^{2}=\omega _{o}^{2}/\omega ^{2}$. Now the equation (64) and the
parameters $S_{bb}$ and $S_{ab}^{2}$ are used to simplify Eq. (58) to the
form 
\begin{equation}
\frac{\omega ^{2}(\Delta x)^{2}}{(\Delta y)^{2}}[(\frac{2m(\Delta y)^{2}}{%
\hslash })^{2}+T_{w}^{2}]=\frac{[1+\tan ^{2}(\omega T)]}{[n_{o}^{2}+\tan
^{2}(\omega T)]}.  \tag{65}
\end{equation}%
Therefore, one obtains these three independent equations (62), (64), and
(65)\ which can be used to determine the three independent parameters $%
n_{o}, $ $\omega _{o}T_{o},$ and $\omega T$ if one is given in advance the
parameters $(\Delta y)^{2},$ $(\Delta x)^{2},$ $T_{w},$ $\omega ,$ and $%
\omega T_{0}.$ First of all, one can solve Eq. (65) to obtain the parameter $%
n_{o}^{2}:$%
\begin{equation}
n_{o}^{2}=\frac{1+\{1-(\frac{\omega ^{2}(\Delta x)^{2}}{(\Delta y)^{2}})[(%
\frac{2m(\Delta y)^{2}}{\hslash })^{2}+T_{w}^{2}]\}\tan ^{2}(\omega T)}{(%
\frac{\omega ^{2}(\Delta x)^{2}}{(\Delta y)^{2}})[(\frac{2m(\Delta y)^{2}}{%
\hslash })^{2}+T_{w}^{2}]}.  \tag{66}
\end{equation}%
Then inserting the parameter $n_{o}^{2}$ into Eq. (64) one obtains 
\begin{equation}
\tan (\omega T)=-\frac{B_{0}}{A_{0}}  \tag{67}
\end{equation}%
where the parameters $A_{0}$ and $B_{0}$ are obtained from the given
parameters $(\Delta y)^{2},$ $(\Delta x)^{2},$ $T_{w},$ $\omega ,$ and $%
\omega T_{0}$ through the equations: 
\begin{equation}
A_{0}=1-(\frac{\omega ^{2}(\Delta x)^{2}}{(\Delta y)^{2}})[(\frac{2m(\Delta
y)^{2}}{\hslash })^{2}+T_{w}^{2}],  \tag{68a}
\end{equation}%
\begin{equation}
B_{0}=\omega T_{0}+\frac{\omega T_{w}(\Delta x)^{2}}{(\Delta y)^{2}}. 
\tag{68b}
\end{equation}%
Now the parameter $\omega T$ can be conveniently determined from Eq. (67)
when the parameters $A_{0}$ and $B_{0}$ are obtained in advance. After the
parameter $\omega T$ is obtained one can determine the parameter $%
n_{o}=\omega _{o}/\omega $ from Eq. (66) and further obtain the parameter $%
\omega _{o}T_{o}$ from Eq. (62) by using the parameters $\omega T$ and $%
n_{o} $. Obviously, given different parameter values $(\Delta y)^{2}$ and $%
T_{w},$ while the other parameters $(\Delta x)^{2},$ $\omega ,$ and $\omega
T_{0}$ are kept constant, one can obtain a different parameter set $\{\omega
_{o},$ $T_{o},$ $\omega ,$ $T\}$. The set of parameters then are used to
generate the two harmonic potential field pulses.

As a summary, in the above discussion one first uses a harmonic potential
field to adjust the parameter $(\Delta x)^{2}$ and then uses a free-particle
motional process and a pair of harmonic potential field pulses to adjust
jointly the complex linewidth of a Gaussian wave-packet motional state. In
these processes the center-of-mass position ($x_{0}=0$) and momentum ($%
p_{0}=0$) of the atomic Gaussian wave-packet motional state always keep
unchanged due to the fact that the atomic motional momentum is zero.
Finally, it can turn out that the inverse unitary propagator of a harmonic
oscillator may be prepared by its unitary propagator. It is well known that
the inverse propagator $U(t,t_{0})^{+}$ of a unitary propagator $U(t,t_{0})$
is just equal to the unitary propagator $U(t_{0},t)$, that is, $%
U(t_{0},t)\equiv U(t,t_{0})^{+}.$ It is known that the unitary propagator of
a harmonic oscillator is given by Eq. (59). If one sets the time interval of
the propagator (59) to be $T=t_{b}-t_{a}=2k\pi /\omega -T^{\prime }$ or $%
\omega T=2k\pi -\omega T^{\prime },$ then the propagator (59) is rewritten as%
\[
G(x_{b},t_{b};x_{a},t_{a})=\sqrt{-\frac{m\omega }{i2\pi \hslash \sin (\omega
T^{\prime })}} 
\]%
\begin{equation}
\times \exp \{-i\frac{m\omega }{2\hslash \sin (\omega T^{\prime })}%
[(x_{b}^{2}+x_{a}^{2})\cos (\omega T^{\prime })-2x_{b}x_{a}]\}.  \tag{69}
\end{equation}%
By comparing Eq. (69) to Eq. (59) one sees that up to a global phase factor
the unitary propagator $U(T)$ (69) is really the inverse propagator $%
U(T^{\prime })^{+}$ of the harmonic oscillator with the time period $%
T^{\prime }$, that is, $U(T)=\exp (i\phi _{0})U(T^{\prime })^{+}$ with a
global phase factor $\exp (i\phi _{0}).$ This means that the inverse unitary
propagator of a harmonic oscillator can be generated from its unitary
propagator (59). One therefore concludes that the Hamiltonian $%
H_{0}=p^{2}/(2m)+m\omega ^{2}x^{2}/2$ of a harmonic oscillator can generate
both the unitary propagator $U(T^{\prime })$ and its inverse propagator $%
U(T^{\prime })$ of the harmonic oscillator. This is completely different
from the case of a free particle. As shown in Eq. (38) and (38a), one needs
to use the inverse unitary propagator of a harmonic oscillator to build up
the state-selective trigger pulse. \newline
\newline
{\large 4. Manipulating a Gaussian wave-packet state by the unitary
propagator of a\ general quadratic Hamiltonian }

In the section it is investigated in detail how the unitary propagator of a
quadratic Hamiltonian (or Lagrangian) affects an atomic Gaussian wave-packet
motional state in an internal-state-independent form. A general method to
manipulate a Gaussian wave-packet state of an atom is to use the unitary
propagator generated by a quadratic Hamiltonian of the atom. A unitary
propagator generated by a quadratic Hamiltonian\ does not change the
Gaussian shape of a Gaussian wave-packet state to any other shape when it
acts on the Gaussian wave-packet state [5, 6, 13, 14]. Generally, a
quadratic Hamiltonian of a quantum system in one dimension may be written as 
\begin{equation}
H(t)=\frac{1}{2m}p^{2}+V(x,t).  \tag{70}
\end{equation}%
Here the generalized quadratic potential operator $V(x,t)$ consists of only
linear and quadratic terms of the center-of-mass coordinate and momentum of
the quantum system, 
\begin{equation}
V(x,t)=\frac{1}{2}b(t)(px+xp)+\frac{1}{2}c(t)x^{2}+d(t)p+f(t)x.  \tag{71}
\end{equation}%
Several typical examples have been given in the previous sections. It is
well known that the potential operator $V(x,t)=0$ for a free particle, $%
V(x,t)=m\omega (t)^{2}x^{2}/2$ for a harmonic oscillator, and $%
V(x,t)=m\omega (t)^{2}x^{2}/2+f(t)x$ for a forced harmonic oscillator. In
the potential operator (71) the linear terms are only responsible for
manipulating the center-of-mass position and momentum of a Gaussian
wave-packet state, while the quadratic terms can be used to control the
complex linewidth of a Gaussian wave-packet state. It is well known that in
the coordinate representation the unitary propagator of a quadratic
Hamiltonian (or Lagrangian) can be exactly obtained by the Feynman path
integration [13, 14, 21]. The time evolution behavior of a quantum system
with a quadratic Hamiltonian has been studied extensively and thoroughly [5,
6, 10, 11, 13, 14, 21, 47, 48, 49]. The unitary propagator of a quantum
system with a quadratic Hamiltonian in one-dimensional coordinate space may
be generally written as [13, 14, 21, 49] 
\[
G(x_{b},t_{b};x_{a},t_{a})=\sqrt{\frac{m}{i2\pi \hslash f_{ab}}}\exp \{i%
\frac{m}{2\hslash }[S_{bb}x_{b}^{2}+S_{ab}2x_{a}x_{b}+S_{aa}x_{a}^{2}]\} 
\]%
\begin{equation}
\times \exp \{+\frac{i}{\hslash }x_{a}Q_{a}(t_{b},t_{a})+\frac{i}{\hslash }%
x_{b}Q_{b}(t_{b},t_{a})\}\exp [i\Theta (t_{b},t_{a})]  \tag{72}
\end{equation}%
where the function $f_{ab}=(-S_{ab})^{-1}$ [49]. Some important and
frequently using unitary propagators which are the special forms of Eq. (72)
have been given in the previous sections: $(i)$ the Hamiltonian of a free
particle is $H=p^{2}/(2m)$ and the unitary propagator is given by Eq. (52a); 
$(ii)$ the Hamiltonian of a harmonic oscillator is given by $H_{0}$ of Eq.
(2)\ and the unitary propagator is given by Eq. (59); $(iii)$ a forced
harmonic oscillator has the Hamiltonian of Eq. (3) and its unitary
propagator is given by [13, 14, 21]%
\[
G_{f}(x_{b},t_{b};x_{a},t_{a})=\exp \{i\Theta
(t_{b},t_{a})\}G(x_{b},t_{b};x_{a},t_{a}) 
\]%
\begin{equation}
\times \exp \{\frac{i}{\hslash }%
[Q_{b}(t_{b},t_{a})x_{b}+Q_{a}(t_{b},t_{a})x_{a}]\}  \tag{73}
\end{equation}%
where $G(x_{b},t_{b};x_{a},t_{a})$ is given by Eq. (59) with the time period 
$T=t_{b}-t_{a}$ and 
\[
Q_{a}(t_{b},t_{a})=-\frac{1}{\sin (\omega T)}\int_{t_{a}}^{t_{b}}f(t)\sin
[\omega (t_{b}-t)]dt, 
\]%
\[
Q_{b}(t_{b},t_{a})=-\frac{1}{\sin (\omega T)}\int_{t_{a}}^{t_{b}}f(t)\sin
[\omega (t-t_{a})]dt, 
\]%
\[
\Theta (t_{b},t_{a})=-\frac{1}{m\omega \hslash \sin (\omega T)}%
\int_{t_{a}}^{t_{b}}\int_{t_{a}}^{t}f(t)f(s)\sin [\omega (t_{b}-t)]\sin
[\omega (s-t_{a})]dsdt. 
\]%
It has been shown that the unitary propagator (73) of a forced harmonic
oscillator can be used to generate a standard coherent state of a harmonic
oscillator [13, 14, 15, 16, 17]. It is known in the section 2 that the
harmonic potential field ($m\omega (t)^{2}x^{2}/2$) and the forced field ($%
f(t)x$) for a harmonic oscillator can be generated by the external driving
electric or magnetic field.

If one names the unitary propagator (72)\ of a quadratic Hamiltonian the
quadratic unitary propagator, then it can turn out that a product of any two
quadratic unitary propagators is still a quadratic unitary propagator. This
property is particularly important as it leads to that a complex quadratic
unitary propagator may be decomposed into a sequence of simple quadratic
unitary propagators. This makes it convenient to implement a complex
quadratic unitary propagator in experiment, since a simple quadratic unitary
propagator can be prepared easily in experiment. The unitary propagator (72)
consists of the quadratic terms $x_{b}^{2},$ $2x_{a}x_{b},$ and $x_{a}^{2}$
and the linear terms $x_{b}$ and $x_{a}$ in addition to the global phase $%
\Theta (t_{b},t_{a}).$ Suppose that there are two quadratic unitary
propagators, each of which contains only quadratic terms $x_{b}^{2},$ $%
2x_{a}x_{b},$ and $x_{a}^{2}.$ Then it can prove that a product of the two
unitary propagators is still a quadratic unitary propagator that contains
only the quadratic terms. This is a direct result of the Lie group generated
by the Lie algebra $su(1,1)$ whose three basis elements may be taken as $%
p^{2},$ $x^{2},$ and $(px+xp)/2.$ On the other hand, a quadratic unitary
propagator which contains linear terms times another quadratic unitary
propagator that contains only quadratic terms will generate a quadratic
unitary propagator that contains linear terms. The quadratic terms of the
unitary propagator (72)\ can manipulate the complex linewidth of a Gaussian
wave-packet state, while the linear terms are used to control only the
center-of-mass position and momentum of the Gaussian wave-packet state. This
will be proven below.

It is known that a standard Gaussian wave-packet state of an atom with mass $%
m$ may be written as [2, 3, 4, 5, 6, 7, 8]%
\[
\Psi _{0}(x,t)=\exp (i\phi _{0})[\frac{(\Delta x)^{2}}{2\pi }]^{1/4}\sqrt{%
\frac{1}{[(\Delta x)^{2}+i(\frac{\hslash T_{0}}{2m})]}}
\]%
\begin{equation}
\times \exp \{-\frac{1}{4}\frac{(x-x_{0})^{2}}{[(\Delta x)^{2}+i(\frac{%
\hslash T_{0}}{2m})]}\}\exp [-ip_{0}x/\hslash ]  \tag{74}
\end{equation}%
where $x_{0}$ and $-p_{0}$ are the center-of-mass position and momentum of
the Gaussian wave-packet state, respectively, $\exp (i\phi _{0})$ is a
global phase factor, and the complex linewidth of the Gaussian wave-packet
state is given by 
\[
W=(\Delta x)^{2}+i(\frac{\hslash T_{0}}{2m}).
\]%
The Gaussian wave-packet state $\Psi _{0}(x,t)$ (74) has the wave-packet
spreading $\varepsilon (T_{0})=\sqrt{2[(\Delta x)^{2}+(\frac{\hslash T_{0}}{%
2m(\Delta x)})^{2}]}.$ The relation between the wave-packet spreading and
the complex linewidth is given by 
\begin{equation}
|W|^{2}=\frac{1}{2}(\Delta x)^{2}\varepsilon (T_{0})^{2}.  \tag{75}
\end{equation}%
The physical meaning for the Gaussian wave-packet state $\Psi _{0}(x,t)$ is
clear: if a free atom is in the Gaussian wave-packet state, then the
Gaussian wave-packet state tells ones that the atom moves along the
direction $x$ in the coordinate axis with the motional velocity $(-p_{0}/m)$%
. A Gaussian wave-packet state is completely described by the three
parameters: the center-of-mass position $x_{0},$ the mean momentum $(-p_{0}),
$ and the complex linewidth $W.$ Now examine the time evolution process of
the Gaussian wave-packet state $\Psi _{0}(x_{a},t_{a})$ (74) under the
action of the unitary propagator $G(x_{b},t_{b};x_{a},t_{a})$ (72) of a
quadratic Hamiltonian. The time evolution process may be calculated by 
\[
\Psi (x_{b},t_{b})=\int dx_{a}G(x_{b},t_{b};x_{a},t_{a})\Psi
_{0}(x_{a},t_{a})
\]%
\[
=\exp (i\phi _{0})[\frac{(\Delta x)^{2}}{2\pi }]^{1/4}\sqrt{\frac{1}{%
[(\Delta x)^{2}+i(\frac{\hslash T_{0}}{2m})]}}\sqrt{\frac{m}{i2\pi \hslash
f_{ab}}}
\]%
\newline
\begin{equation}
\times \int_{-\infty }^{\infty }dx_{a}\{\exp \{iS_{c}/\hslash \}\exp \{-%
\frac{1}{4}\frac{(x_{a}-x_{0})^{2}}{[(\Delta x)^{2}+i(\frac{\hslash T_{0}}{2m%
})]}\}\exp \{-ip_{0}x_{a}/\hslash \}\}  \tag{76a}
\end{equation}%
where the action $S_{c}$ can be found from the propagator (72), 
\[
S_{c}=\frac{1}{2}%
m[S_{bb}x_{b}^{2}+S_{ab}2x_{a}x_{b}+S_{aa}x_{a}^{2}]+x_{a}Q_{a}(t_{b},t_{a})+x_{b}Q_{b}(t_{b},t_{a})+\hslash \Theta (t_{b},t_{a}).
\]%
One can write the final state $\Psi (x_{b},t_{b})$ as, with the help of the
Gaussian integral (50),%
\[
\Psi (x_{b},t_{b})=\exp (i\phi _{0})\exp [i\Theta (t_{b},t_{a})][\frac{%
(\Delta x)^{2}}{2\pi }]^{1/4}\sqrt{\frac{1}{[(\Delta x)^{2}+i(\frac{\hslash
T_{0}}{2m})]}}\sqrt{\frac{m}{i2\pi \hslash f_{ab}}}
\]%
\begin{equation}
\times \exp \{-\frac{1}{4}\frac{x_{0}^{2}}{[(\Delta x)^{2}+i(\frac{\hslash
T_{0}}{2m})]}\}\exp \{\frac{im}{2\hslash }S_{bb}x_{b}^{2}\}\exp \{\frac{i}{%
\hslash }x_{b}Q_{b}(t_{b},t_{a})\}\sqrt{\frac{\pi }{a}}\exp (\frac{b^{2}}{4a}%
),  \tag{76b}
\end{equation}%
where the parameters $a$ and $b$ are given by 
\begin{equation}
a=-i\frac{m}{2\hslash }S_{aa}+\frac{1}{4}\frac{1}{[(\Delta x)^{2}+i(\frac{%
\hslash T_{0}}{2m})]}  \tag{77a}
\end{equation}%
and 
\begin{equation}
b=\frac{im}{\hslash }S_{ab}x_{b}+\frac{i}{\hslash }%
[Q_{a}(t_{b},t_{a})-p_{0}]+\frac{1}{2}\frac{x_{0}}{[(\Delta x)^{2}+i(\frac{%
\hslash T_{0}}{2m})]}.  \tag{77b}
\end{equation}%
\newline
Since $b^{2}/(4a)$ is quadratic with the coordinate $x_{b},$ the wave
function $\Psi (x_{b},t_{b})$ of Eq. (76b) can be written as 
\[
\Psi (x_{b},t_{b})=\exp (i\phi _{0})\exp [i\Theta (t_{b},t_{a})][\frac{%
(\Delta x)^{2}}{2\pi }]^{1/4}
\]%
\begin{equation}
\times \sqrt{\frac{1}{f_{ab}}\frac{1}{S_{aa}(\Delta x)^{2}+i(\frac{\hslash }{%
2m})[1+T_{0}S_{aa}]}}\exp \{Ax_{b}^{2}+Bx_{b}+C)\}  \tag{78}
\end{equation}%
where the complex coefficients A, B, and C are determined through 
\begin{equation}
Ax_{b}^{2}+Bx_{b}+C=\frac{b^{2}}{4a}+\frac{im}{2\hslash }S_{bb}x_{b}^{2}+%
\frac{i}{\hslash }x_{b}Q_{b}(t_{b},t_{a})-\frac{1}{4}\frac{x_{0}^{2}}{%
[(\Delta x)^{2}+i\frac{\hslash T_{0}}{2m}]}.  \tag{79}
\end{equation}%
\newline
By substituting the parameter $a$ of Eq. (77a)\ and $b$ of Eq. (77b) in the
equation (79) one can obtain explicitly the complex coefficients A, B, and
C. Furthermore one can find, by a complex calculation, 
\[
\exp \{\func{Re}(A)x_{b}^{2}+\func{Re}(B)x_{b}+\func{Re}(C)\}
\]%
\[
=\exp \{-\frac{1}{4}S_{ab}^{2}\frac{\{x_{b}+S_{ab}^{-1}[S_{aa}x_{0}+(\frac{%
Q_{a}(t_{b},t_{a})-p_{0}}{m})]\}^{2}}{S_{aa}^{2}(\Delta x)^{2}+(\frac{%
\hslash }{2m(\Delta x)})^{2}[1+S_{aa}T_{0}]^{2}}\}.
\]%
Therefore, the wave function $\Psi (x_{b},t_{b})$ of Eq. (78) can be further
written as 
\[
\Psi (x_{b},t_{b})=\exp (i\phi _{0})\exp [i\Theta (t_{b},t_{a})]\exp \{i%
\func{Im}(C)\}
\]%
\[
\times \lbrack \frac{(\Delta x)^{2}}{2\pi }]^{1/4}\sqrt{\frac{(-S_{ab})}{%
S_{aa}(\Delta x)^{2}+i(\frac{\hslash }{2m})[1+T_{0}S_{aa}]}}
\]%
\newline
\[
\times \exp \{-\frac{1}{4}S_{ab}^{2}\frac{\{x_{b}+S_{ab}^{-1}[S_{aa}x_{0}+(%
\frac{Q_{a}(t_{b},t_{a})-p_{0}}{m})]\}^{2}}{S_{aa}^{2}(\Delta x)^{2}+(\frac{%
\hslash }{2m(\Delta x)})^{2}[1+S_{aa}T_{0}]^{2}}\}
\]%
\begin{equation}
\times \exp \{i[\func{Im}(A)x_{b}^{2}+\func{Im}(B)x_{b}]\}.  \tag{80}
\end{equation}%
The wave function $\Psi (x_{b},t_{b})$ of Eq. (80)\ is indeed a Gaussian
wave-packet state. This can be seen more clearly from its probability
density $|\Psi (x_{b},t_{b})|^{2}$ which is a standard Gaussian function.
Thus, the Gaussian wave-packet state $\Psi (x_{b},t_{b})$ (80)\ has the
center-of-mass position: 
\[
x_{c}=-S_{ab}^{-1}[S_{aa}x_{0}+(\frac{Q_{a}(t_{b},t_{a})-p_{0}}{m})]
\]%
and the wave-packet spreading: 
\[
\varepsilon =\sqrt{2S_{ab}^{-2}\{S_{aa}^{2}(\Delta x)^{2}+(\frac{\hslash }{%
2m(\Delta x)})^{2}[1+T_{0}S_{aa}]^{2}\}}.
\]%
\newline
The imaginary parts of the coefficients A, B, and C in the state $\Psi
(x_{b},t_{b})$ (80)\ are given by 
\begin{equation}
\func{Im}(A)=(\frac{m}{2\hslash })\frac{%
S_{bb}(1+S_{aa}T_{0})+Z(T_{0})(S_{bb}S_{aa}-S_{ab}^{2})}{\delta
^{2}S_{aa}^{2}+[1+S_{aa}T_{0}]^{2}},  \tag{81a}
\end{equation}%
\begin{equation}
\func{Im}(B)=(\frac{m}{\hslash })\frac{%
(v_{b}+S_{ab}x_{0})(1+S_{aa}T_{0})+Z(T_{0})(v_{b}S_{aa}-v_{a}S_{ab})}{%
[1+S_{aa}T_{0}]^{2}+\delta ^{2}S_{aa}^{2}},  \tag{81b}
\end{equation}%
\begin{equation}
\func{Im}(C)=(\frac{m}{2\hslash })\frac{%
(x_{0}^{2}S_{aa}+2x_{0}v_{a}-T_{0}v_{a}^{2})[1+S_{aa}T_{0}]-\delta
^{2}S_{aa}v_{a}^{2}}{\delta ^{2}S_{aa}^{2}+[1+S_{aa}T_{0}]^{2}},  \tag{81c}
\end{equation}%
where $Z(T_{0})=[S_{aa}\delta ^{2}+T_{0}(1+S_{aa}T_{0})]$ and the parameters 
$\delta ,$ $v_{a},$ and $v_{b}$ are defined by%
\[
\delta =(\frac{2m(\Delta x)^{2}}{\hslash }),\text{ \ }v_{a}=\frac{1}{M}%
[Q_{a}(t_{b},t_{a})-p_{0}],\text{ \ }v_{b}=\frac{1}{M}Q_{b}(t_{b},t_{a}).
\]%
The state $\Psi (x_{b},t_{b})$ (80)\ can be further written as a standard
Gaussian wave-packet state: 
\[
\Psi (x,t)=\exp (i\varphi _{0})[\frac{(\Delta y)^{2}}{2\pi }]^{1/4}\sqrt{%
\frac{1}{[(\Delta y)^{2}+i(\frac{\hslash T_{w}}{2m})]}}
\]%
\newline
\begin{equation}
\times \exp \{-\frac{1}{4}\frac{(x-x_{c})^{2}}{[(\Delta y)^{2}+i(\frac{%
\hslash T_{w}}{2m})]}\}\exp [-ipx/\hslash ]  \tag{82}
\end{equation}%
where $\exp (i\varphi _{0})$ is a global phase factor, the mean momentum $%
(-p)$ is given through%
\begin{equation}
p/\hslash =-\func{Im}(B)-2x_{c}\func{Im}(A),  \tag{83}
\end{equation}%
\newline
the real part of the complex linewidth is 
\begin{equation}
(\Delta y)^{2}=\frac{S_{ab}^{2}\{S_{aa}^{2}(\Delta x)^{2}+(\frac{\hslash }{%
2m(\Delta x)})^{2}[1+S_{aa}T_{0}]^{2}\}}{S_{ab}^{4}+16(\func{Im}%
(A))^{2}\{S_{aa}^{2}(\Delta x)^{2}+(\frac{\hslash }{2m(\Delta x)}%
)^{2}[1+S_{aa}T_{0}]^{2}\}^{2}},  \tag{84}
\end{equation}%
and the imaginary part equals 
\begin{equation}
(\frac{\hslash T_{w}}{2m})=\frac{4\func{Im}(A)\{S_{aa}^{2}(\Delta x)^{2}+(%
\frac{\hslash }{2m(\Delta x)})^{2}[1+S_{aa}T_{0}]^{2}\}^{2}}{S_{ab}^{4}+16(%
\func{Im}(A))^{2}\{S_{aa}^{2}(\Delta x)^{2}+(\frac{\hslash }{2m(\Delta x)}%
)^{2}[1+S_{aa}T_{0}]^{2}\}^{2}}.  \tag{85}
\end{equation}%
Here the imaginary part of the complex linewidth can be positive or
negative, which is dependent on the term $\func{Im}(A).$ It is known from
Eq. (81a) that the term $\func{Im}(A)$ is dependent on only the parameters $%
S_{bb},$ $S_{ab},$ and $S_{aa}$ of the quadratic terms in the unitary
propagator (72) of the quadratic Hamiltonian (70) but independent of those
parameters of the linear terms. Then the real ($(\Delta y)^{2}$) and the
imaginary ($T_{w}$) part of the complex linewidth are dependent on only the
quadratic terms but not the linear terms of the unitary propagator (72). One
therefore concludes that the complex linewidth of a Gaussian wave-packet
state can be adjusted only by the quadratic terms in the unitary propagator
(72). This is a convenient method to manipulate the complex linewidth of a
Gaussian wave-packet state by adjusting only the quadratic terms of the
unitary propagator (72). In the quadratic Hamiltonian (70) the linear terms $%
d(t)p$ and $f(t)x$ do not have a contribution to the quadratic terms of the
unitary propagator (72) [13, 14, 21, 49]. Then one can use only the
quadratic operator terms of the Hamiltonian (70) to adjust the quadratic
terms of the unitary propagator (72).\ Consequently one can manipulate at
will the complex linewidth of a Gaussian wave-packet state by the quadratic
operator terms of the quadratic Hamiltonian (70). On the other hand, it is
known from Eq. (81b) that the term $\func{Im}(B)$ is dependent on the
parameters $Q_{a}(t_{b},t_{a})$ and $Q_{b}(t_{b},t_{a})$ of the linear terms
of the unitary propagator (72). Then the center-of-mass position $x_{c}$ and
momentum $(-p)$ of the Gaussian wave-packet state (82) are dependent on the
parameters $Q_{a}(t_{b},t_{a})$ and/or $Q_{b}(t_{b},t_{a})$, although they
are also dependent on those parameters of the quadratic terms. One therefore
can manipulate the center-of-mass position and momentum of a Gaussian
wave-packet state through the linear terms of the unitary propagator (72) or
more conveniently through the linear terms of the quadratic Hamiltonian
(70). A typical example is that in order to generate a standard coherent
state of harmonic oscillator with a higher motional energy one may use the
linear term $f(t)x$ [13, 14, 15, 16, 17, 21, 49] that may be generated by
the external driving field, while one may use the time-dependent and
frequency-varying harmonic potential field to adjust the complex linewidth
of a Gaussian wave-packet state, as shown in the previous section 3. \newline
\newline
{\large 5. Discussion}

The state-selective trigger pulse has the two basic properties. One of which
is that the state-selective trigger pulse can have a real effect on the atom
only when the atom is in some given internal states. Another is that the
state-selective trigger pulse does not change Gaussian shape of an atomic
Gaussian wave-packet motional state to any other shape. The former property
is inherent and the last one is due to the fact that a Gaussian wave-packet
motional state for a single atom is simple and easy to be manipulated and
controlled in time and space. An internal-state-dependent selective
excitation process of an atomic system is generally closely related to
manipulation and control in time and space of the atomic internal electronic
(or spin) motion, the atomic center-of-mass motion, and the coupling between
the internal and the center-of-mass motion, although an inhomogeneous
external magnetic field could also generate an internal-state-dependent
force exerted on a spin and hence could be used to generate in an
internal-state selective form a coherent state of the spin in the harmonic
potential field. Therefore, a general construction for the state-selective
trigger pulse is generally involved in using the electromagnetic field
pulses, i.e., the laser light pulses, to create the interaction between the
atomic internal states and center-of-mass motional states and realize the
coupling between the center-of-mass and the internal motion of the atom. A
state-selective trigger pulse transfers one atomic Gaussian wave-packet
motional state to another with the help of the atomic internal states. This
point is the same as those of the unitary decelerating and accelerating
processes [8]. On the other hand, there are many other space-dependent
processes in quantum information science. Typical examples include the
quantum communication [52] and the construction of quantum gates by the
short-range interactions such as the conditional collision interaction [53].
These space-dependent processes generally use the particle (photon or atom)
transport process to realize the quantum state transfer of the atomic
internal states or the photon polarization states in space and implement the
quantum gate operations of the internal-state quantum bits. Here motional
states are usually considered as carrier of the quantum information
transport. These processes generally emphasize realization of the
internal-state transfer in space (i.e., quantum information transfer) or the
quantum gate operations of the atomic internal-state quantum bits with the
help of the motional states of the particles instead of the unitary
manipulation of the motional states themselves. In this sense these
space-dependent processes are different from the state-selective excitation
process of the trigger pulse and the unitary decelerating and accelerating
processes. In the reversible and unitary halting protocol [1] the quantum
program converts the difference of the atomic internal states (the initial
functional states) into the difference between the wave-packet motional
states of the halting-qubit atom in space. Thus, one has to manipulate in
time and space the atomic wave-packet motional states, in order that the
reversible and unitary halting protocol is state-insensitive. Note that here
the atomic wave-packet motional states are not used as quantum bits and the
atomic internal states still act as the halting quantum bit.

Manipulating unitarily in space a wave-packet motional state of an atom or a
superposition of motional states of the atom is generally difficult in
experiment with respect to manipulating a purely time-dependent quantum
state. However, it is of crucial importance to realize both the reversible
and unitary state-insensitive halting protocol and the efficient quantum
search process. The quantum control process that simulates the reversible
and unitary halting protocol contains the conventional particle transport
process such as the free-particle motion, but the more important is that it
also contains the coherent-state selective excitation process of the trigger
pulse and the unitary decelerating and accelerating processes that are
different from the conventional particle transport process. The conventional
transport process usually need not require the cooperation of the
center-of-mass and the internal motion of the atom, while these unitary
processes need to manipulate not only the atomic internal and center-of-mass
motions in space but also their coupling. Unitary manipulation for the
atomic internal states which act as quantum bits is an important research
area in quantum computation in the past decade, and it is usually more
convenient than for atomic wave-packet motional states. One of the main
reasons why it is generally difficult to manipulate at will an atomic
wave-packet motional state in space is that the unitary dynamical process is
generally complicated for an atom in a general potential field. In few
simplest potential fields such as a harmonic potential field the quantum
dynamical behavior of an atom can be completely understood, while a complete
knowledge for an atom system and its quantum dynamics is the basis to
manipulate at will the atomic motional states in space. Thus, at present the
unitary manipulation in time and space of motional states of the
halting-qubit atom in the quantum control process has to be restricted to a
simplest case such as the atom in a Gaussian wave-packet motional state and
in a harmonic potential field. In contrast to the preparation of quantum
gate operations, here the unitary manipulation in time and space of the
atomic motional states becomes the main research area.

The quantum dynamics of a harmonic oscillator and a Gaussian wave-packet
state have been studied extensively and thoroughly in quantum mechanics. A
Gaussian wave-packet state is one of the simplest quantum wave-packet states
that can be manipulated and controlled in time and space easily and
precisely. In the previous paper [8] it has been shown that the unitary
decelerating and accelerating processes based on the STIRAP method can
transfer one Gaussian wave-packet motional state of a free atom to another
in the ideal or near ideal adiabatic condition and can manipulate the
center-of-mass position and momentum of a Gaussian wave-packet motional
state. The advantage of the manipulation is that the space-selective
manipulation can be carried out easily, since the manipulation uses the
STIRAP laser light pulse sequence, while laser light makes it easy to
perform the space-selective and internal-state-selective operations of an
atom. However, the shortcoming of the manipulation is that the complex
linewidth of a Gaussian wave-packet motional state\ can not be manipulated
at will by the STIRAP method. Now in this paper several methods have been
developed to manipulate the complex linewidth of a Gaussian wave-packet
motional state of an atom. Their basic starting point is to apply a
quadratic potential field to the atom. These results in the previous [8] and
the present paper show that a Gaussian wave-packet motional state of an atom
can be manipulated at will in experiment. It can be predicted from these
results that there is no longer unsurpassable obstacle in theory for a
quantum computer to solve the unsorted quantum search problem in polynomial
time. \newline
\newline
{\large References }\newline
1. X. Miao, \textit{The basic principles to construct a generalized
state-locking pulse field and simulate efficiently the reversible and
unitary halting protocol of a universal quantum computer},
http://arxiv.org/abs/quant-ph/0607144 (2006) \newline
2. L. I. Schiff, \textit{Quantum mechanics}, 3rd, McGraw-Hill book company,
New York, 1968\newline
3. M. L. Goldberger and K. M. Watson, \textit{Collision theory}, Chapt. 3,
Wiley, New York, 1964\newline
4.\ R. G. Newton, \textit{Scattering theory of waves and particles}, Chapt.
6, McGraw-Hill, New York, 1966.\newline
5.\ E. J. Heller, \textit{Time-dependent approach to semiclassical dynamics}%
, J. Chem. Phys. 62, 1544 (1975) \newline
6. D. Huber and E. J. Heller, \textit{Generalized Gaussian wave packet
dynamics}, J. Chem. Phys. 87, 5302 (1987)\newline
7. R. G. Littlejohn, \textit{The semiclassical evolution of wave packets},
Phys. Rep. 138, 193 (1986)\newline
8. X. Miao, \textit{The STIRAP-based unitary decelerating and accelerating
processes of a single free atom}, http://arxiv.org/abs/quant-ph/0707.0063
(2007)\newline
9. K. Bergmann, H. Theuer, and B. W. Shore, \textit{Coherent population
transfer among quantum states of atoms and molecules}, Rev. Mod. Phys. 70,
1003 (1998)\newline
10. R. J. Glauber, \textit{Coherent and incoherent states of the radiation
field}, Phys. Rev. 131, 2766 (1963)\newline
11. (a) R. F. Bishop and A. Vourdas, \textit{Generalised coherent states and
Bogoliubov transformations}, J. Phys. A 19, 2525 (1986); (b) J. N.
Hollenhorst, \textit{Quantum limits on resonant-mass gravitational-radiation
detectors}, Phys. Rev. D 19, 1669 (1979)\newline
12. X. Miao, \textit{Quantum search processes in the cyclic group state
spaces}, http:// arxiv.org/abs/quant-ph/0507236 (2005)\newline
13. L. S. Schulman, \textit{Techniques and applications of path integration}%
, Dover, New York, 2005\newline
14. R. P. Feynman, \textit{Space-time approach to non-relativistic quantum
mechanics}, Rev. Mod. Phys. 20, 367 (1948); R. P. Feynmann and A. R. Hibbs, 
\textit{Quantum mechanics and path integrals}, McGraw-Hill, New York, 1965 
\newline
15. K. Husimi, \textit{Miscellanea in elementary quantum mechanics II},
Prog. Theor. Phys. 9, 381 (1953)\newline
16. E. H. Kerner, \textit{Note on the forced and damped oscillator in
quantum mechanics}, Can. J. Phys. 36, 371 (1957)\newline
17. P. Carruthers and M. M. Nieto, \textit{Coherent states and the forced
quantum oscillator}, Am. J. Phys. 33, 537 (1965)\newline
18. (a)\ D. M. Meekhof, C. Monroe, B. E. King, W. M. Itano, and D. J.
Wineland, \textit{Generation of nonclassical motional states of a trapped
atom}, Phys. Rev. Lett. 76, 1796 (1996); Erratum, 77, 2346 (1996); (b) C.
Monroe, D. M. Meekhof, B. E. King, and D. J. Wineland, \textit{A }$^{\prime
\prime }$\textit{Schr\"{o}dinger cat}$^{\prime \prime }$\textit{\
superposition state of an atom,} Science 272, 1131 (1996) \newline
19. D. J. Wineland, C. Monroe, W. M. Itano, D. Leibfried, B. E. King, D. M.
Meekhof, \textit{Experimental issues in coherent quantum-state manipulation
of trapped atomic ions}, J. Res. NIST, 103, 259 (1998) \newline
20. D. Leibfried, R. Blatt, C. Monroe, and D. Wineland, \textit{Quantum
dynamics of single trapped ion}, Rev. Mod. Phys. 75, 281 (2003)\newline
21. C. Grosche and F. Steiner, \textit{Handbook of Feynman path integrals},
Springer, Berlin, 1998 \newline
22. (a)\ S. Chu, \textit{Nobel Lecture: The manipulation of neutral particles%
}, Rev. Mod. Phys. 70, 685 (1998)

(b) C. N. Cohen-Tannoudji, \textit{Nobel Lecture: Manipulating atoms with
photons}, Rev. Mod. Phys. 70, 707 (1998)

(c)\ W. D. Phillips, \textit{Nobel Lecture: Laser cooling and trapping of
neutral atoms}, Rev. Mod. Phys. 70, 721 (1998)\newline
23. J. I. Cirac and P. Zoller, \textit{Quantum computations with cold
trapped ions}, Phys. Rev. Lett. 74, 4091 (1995)\newline
24. C. A. Blockley, D. F. Walls, and H. Risken, \textit{Quantum collapses
and revivals in a quantized trap}, Europhys. Lett. 17, 509 (1992)\newline
25. (a)\ J. I. Cirac, A. S. Parkins, R. Blatt, and P. Zoller, \textit{"Dark"
squeezed states of the motion of a trapped ion}, Phys. Rev. Lett. 70, 556
(1993); (b) J. I. Cirac, R. Blatt, and P. Zoller, \textit{Nonclassical
states of motion in a three-dimensional ion trap by addiabatic passage},
Phys. Rev. A 49, R3174 (1994) \newline
26. W. Vogel and R. L. de Matos Filho, \textit{Nonlinear Jaynes-Cummings
dynamics of a trapped ion}, Phys. Rev. A 52, 4214 (1995); \newline
27. C. Monroe, D. M. Meekhof, B. E. King, S. R. Jefferts, W. M. Itano, D. J.
Wineland, and P. Gould, \textit{Ressolved-sideband Raman cooling of a bound
atom to the 3D zero-point energy}, Phys. Rev. Lett. 75, 4011 (1995)\newline
28. P. Marte, P. Zoller, and J. L. Hall, \textit{Coherent atomic mirrors and
beam splitters by adiabatic passage in multilevel systems}, Phys. Rev. A 44,
R4118 (1991)\newline
29. M. Weitz, B. C. Young, and S. Chu, \textit{Atom manipulation based on
delayed laser pulses in three- and four-level systems: light shifts and
transfer efficiencies}, Phys. Rev. A 50, 2438 (1994) \newline
30. J. L. S\o rensen, D. Moller, T. Iversen, J. B. Thomsen, F. Jensen, P.
Staanum, D. Voigt, and M. Drewsen, \textit{Efficient coherent internal state
transfer in trapped ions using stimulated Raman adiabatic passage},
http://arxiv.org/abs/quant-ph /0608089 (2006) \newline
31. S. Wallentowitz and W. Vogel, \textit{Quantum-mechanical counterpart of
nonlinear optics}, Phys. Rev. A 55, 4438 (1997)\newline
32. G. S. Agarwal and J. Banerji, \textit{Quantum evolution of classical
nonlinear eigenmode in parametric interaction and realization in traps with
ions}, Phys. Rev. A 55, R4007 (1997) \newline
33. J. Steinbach, J. Twamley, and P. L. Knight, \textit{Engineering two-mode
interactions in ion traps}, Phys. Rev. A 56, 4815 (1997) \newline
34. L. Allen and J. H. Eberly, \textit{Optical resonance and two-level atoms}%
, Dover, New York, 1987\newline
35. E. T. Jaynes and F. W. Cummings, \textit{Comparison of quantum and
semiclassical radiation theories with application to the beam maser}, Proc.
Inst. Electr. Eng. 51, 89 (1963)\newline
36. See, for example, (a)\ J. Oreg, F. T. Hioe, and J. H. Eberly, \textit{%
Adiabatic following in multilevel systems}, Phys. Rev. A 29, 690 (1984); (b)
E. Brion, L. H. Pedersen, and K. Molmer, \textit{Adiabatic elimination in a
Lambda system}, http://arxiv.org/abs/quant-ph/0610056 (2006)\newline
37. M. Matti Maricq, \textit{Application of average Hamiltonian theory to
the NMR of solids}, Phys. Rev. B 25, 6622 (1982)\newline
38. C. K. Law and J. H. Eberly, \textit{Arbitrary control of a quantum
electromagnetic field}, Phys. Rev. Lett. 76, 1055 (1996)\newline
39. W. Magnus, \textit{On the exponential solution of differential equations
for a linear operator}, Commun. Pure Appl. Math. 7, 649 (1954) \newline
40. H. F. Trotter, \textit{On the product of semigroups of operators}, Proc.
Am. Math. Soc. 10, 545 (1959)\newline
41. R. M. Wilcox, \textit{Exponential operators and parameter
differentiation in quantum physics}, J. Math. Phys. 8, 962 (1967) \newline
42. M. Suzuki, \textit{Fractal decomposition of exponential operators with
applications to many-body theories and Monte Carlo simulations}, Phys. Lett.
A 146, 319 (1990); \textit{General theory of higher-order decomposition of
exponential operators and symplectic integrators}, Phys. Lett. A 165, 387
(1992) \newline
43. (a)\ H. Yoshida, \textit{Construction of higher order symplectic
integrators}, Phys. Lett. A 150, 262 (1990); (b)\ W. S. Zhu and X. S. Zhao, 
\textit{Numerical quantum propagation with time-dependent Hamiltonian}, J.
Chem. Phys. 105, 9536 (1996)\newline
44. A. S$\phi $rensen and K. M$\phi $lmer, \textit{Quantum computation with
ions in thermal motion}, Phys. Rev. Lett. 82, 1971 (1999)\newline
45. J. Milburn, \textit{Simulating nonlinear spin models in an ion trap},
http://arxiv.org /abs/quant-ph/9908037 (1999)\newline
46. J. F. Poyatos, J. I. Cirac, and P. Zoller, \textit{Quantum gates with
"hot" trapped ions}, Phys. Rev. Lett. 81, 1322 (1998)\newline
47. P. A. Horvathy, \textit{Extended Feynman formula for harmonic oscillator}%
, Int. J. Theor. Phys. 18, 245 (1979)\newline
48. I. H. Duru, \textit{Feynman}$^{\prime }$\textit{s formula for a harmonic
oscillator}, Int. J. Theor. Phys. 23, 567 (1984)\newline
49. (a) M. M. Mizrahi, \textit{Phase space path integrals, without limiting
procedures}, J. Math. Phys. 19, 298 (1978); (b) D. C. Khandekar and S. V.
Lawande, \textit{Exact solution of a time-dependent quantal harmonic
oscillator with damping and a perturbative force}, J. Math. Phys. 20, 1870
(1979); (c) B. K. Cheng, \textit{The propagator of the time-dependent forced
harmonic oscillator with time-dependent damping}, J. Math. Phys. 27, 217
(1986); (d) J. T. Marshall and J. T. Pell, \textit{Path-integral evaluation
of the space-time propagator for quadratic Hamiltonian systems}, J. Math.
Phys. 20, 1297 (1979) \newline
50. A. Abragam, \textit{Principles of nuclear magnetism}, Oxford University
Press, London, 1961\newline
51. X. Miao, \textit{Multiple-quantum operator algebra spaces and
description for the unitary time evolution of multilevel spin systems},
Molec. Phys. 98, 625 (2000) \newline
52. C.\ H. Bennett and G. Brassard, \textit{Quantum cryptography : public
key distribution and coin tossing, Proceedings of IEEE international
conference on computers, systems, and signal processing,} pp. 175--179,
IEEE, New York, 1984\newline
53. T. Calarco, E. A. Hinds, D. Jaksch, J. Schmiedmayer, J. I. Cirac, and P.
Zoller, \textit{Quantum gates with neutral atoms: Controlling collisional
interactions in time dependent traps}, Phys. Rev. A 61, 022304 (2000)\newline
54. (a) U. Haeberlen and J. Waugh, \textit{Coherent averaging effects in
magnetic resonance}, Phys. Rev. 175, 453 (1968); (b) U. Haeberlen, \textit{%
High resolution NMR in solids}, Adv. Magn. Reson. Suppl. 1, 1976

\end{document}